\documentclass{article}
\usepackage{fullpage}

\usepackage{relsize}
\usepackage{multirow} 
\usepackage{amsmath, amsthm, amsfonts,amssymb}
\usepackage{bm,bbm}
\usepackage{graphicx,float,wrapfig}
\usepackage{subfigure}
\DeclareMathAlphabet{\mathcal}{OMS}{cmsy}{m}{n}

\PassOptionsToPackage{colorlinks = true}{hyperref}
\usepackage{hyperref}
\usepackage{tabu}
\usepackage{url} 
\usepackage[colorlinks = true]{hyperref}
\usepackage{cite}
\usepackage{algpseudocode}
\usepackage{algorithm}
\usepackage{graphicx} 
\usepackage{cancel}
\usepackage{adjustbox}

\newtheorem{proposition}{Proposition}

\newtheorem{theorem}{Theorem}

\newcommand{\cqfd}{\hfill\rule{2mm}{2mm}\medbreak\indent}

\newcommand{\e}{\begin{equation}}
\newcommand{\ee}{\end{equation}}
\newcommand{\eqn}{\begin{eqnarray}}
\newcommand{\eeqn}{\end{eqnarray}}
\newcommand{\MAT}{\left[ \begin{array}}
\newcommand{\mat}{\end{array} \right]}

\newtheorem{Remark}{Remark}[section]

\usepackage{xcolor}  
\usepackage{soul}

\setstcolor{red}

\newcommand{\vct}[1]{\boldsymbol{#1}}
\newcommand{\mtx}[1]{\boldsymbol{#1}}

\newcommand{\ve}{\vct{e}}

\newcommand{\vr}{\vct{r}}
\newcommand{\vs}{\vct{s}}

\newcommand{\vv}{\vct{v}}
\newcommand{\vw}{\vct{w}}
\newcommand{\vx}{\vct{x}}

\newcommand{\vz}{\vct{z}}
\newcommand{\valpha}{\vct{\alpha}}
\newcommand{\vbeta}{\vct{\beta}}

\newcommand{\vxi}{\vct{\xi}}

\newcommand{\vzero}{\vct{0}}

\newcommand{\mA}{\mtx{A}}

\newcommand{\mD}{\mtx{D}}

\newcommand{\mG}{\mtx{G}}

\newcommand{\mI}{\mtx{I}}

\newcommand{\mQ}{\mtx{Q}}
\newcommand{\mR}{\mtx{R}}

\newcommand{\mT}{\mtx{T}}
\newcommand{\mU}{\mtx{U}}
\newcommand{\mV}{\mtx{V}}
\newcommand{\mW}{\mtx{W}}

\newcommand{\mOmega}{\mtx{\Omega}}

\newcommand{\mSigma}{\mtx{\Sigma}}

\newcommand{\mGamma}{\mtx{\Gamma}}

\newcommand{\mzero}{{\bf 0}}

\usepackage{bm}   

\usepackage{color}
\newcommand{\SL}[1]{{\color{black} #1}} 

\begin{document}

 
\title{On A Class of Greedy Sparse Recovery Algorithms}

\author{Gang Li\thanks{School of Automation and Electrical Engineering, Zhejiang University of Science \& Technology, Hangzhou, Zhejiang, 310023, P.R. China. Email: ieligang@zjut.edu.cn}, 
Qiuwei Li\thanks{DAMO Academy, Alibaba Group US, Bellevue, Washington, 98004, USA. Email: liqiuweiss@gmail.com}, 
Shuang Li\thanks{Department of Electrical and Computer Engineering, Iowa State University, Ames, Iowa, 50014, USA. Email: lishuang@iastate.edu. ORCID: 0000-0003-3729-7496}, 
and Wu Angela Li\thanks{Department of Computer Science, Rice University, Houston, Texas, 77005, USA. Email: awl@rice.edu}}

\maketitle

\begin{abstract}

Sparse signal recovery deals with finding the sparsest solution of an under-determined linear system $\vx = \mQ\vs$. In this paper, we propose a novel greedy approach to addressing the challenges from such a problem. Such an approach is based on a characterization of solutions to the system, which allows us to work on the sparse recovery in the $\vs$-space directly with a given measure. With $l_2$-based measure, an orthogonal matching pursuit (OMP)-type algorithm is proposed, which significantly outperforms the classical OMP algorithm in terms of recovery accuracy while maintaining comparable computational complexity. An $l_1$-based algorithm, denoted as $\text{Alg}_{GL1}$,
is derived. Such an algorithm significantly outperforms the classical basis pursuit (BP) algorithm. Combining with the CoSaMP-strategy for selecting atoms, a class of high performance greedy algorithms is also derived. Extensive numerical simulations on both synthetic and image data are carried out, with which the superior performance of our proposed algorithms is demonstrated in terms of sparse recovery accuracy and robustness against numerical instability of the system matrix $\mQ$ and disturbance in the measurement $\vx$.
\end{abstract}


\section{Introduction} \label{sec-1}

The sparse recovery problem, also known as the {\it sparse linear inverse} problem, has found extensive applications in many areas such as data science~\cite{shikhman2021sparse,osher2019sparse}, signal processing~\cite{stankovic2019tutorial,li2019atomic,davies2012rank,durgin2019jointly}, communications engineering~\cite{jin2011limits,zhang2023near,li2020adaptive}, structural health monitoring~\cite{huang2016bayesian,li2018atomic,tang2021group}, hyperspectral imaging~\cite{golbabaee2012hyperspectral,arad2016sparse,durgin2019fast}, and deep learning~\cite{lecun2015deep,goodfellow2016deep}. It plays a crucial role in the framework of compressed sensing (CS)~\cite{donoho2006compressed,candes2008introduction,duarte2011structured,zhu2018collaborative,vershynin2012compressed,Foucart2013mathematical}, and in deep learning for feature selection and designing efficient convolutional neural networks~\cite{gui2016feature,xu2023sparse,krizhevsky2012imagenet,liu2015sparse,friedman2010regularization,liu2020structured}.  

Consider the following  popularly used synthetic model:
\e \vx = \mQ \vs,\label{Synthetic-model}\ee
where $\mQ \in \mathbb R^{N\times L}$ with $N < L$, and $\vs \in \mathbb R^{L\times 1}$ is the representation of the measurement $\vx$ in $\mQ$. The associated sparse recovery problem regards for a given $(\vx, \mQ)$, finding such a solution of (\ref{Synthetic-model}), denoted as $\vs^\star$, which has fewer nonzero entries than any other solution. Mathematically, it is formulated as


\e \vs^\star\triangleq \arg\min_{\vs}\|\vs\|_0~~~~\text{s.t.}~~\vx = \mQ\vs, \label{f-1}\ee
where $\|\vv\|_0$, often denoted as $l_0$ norm though it is not a norm in a strict sense, represents the number of nonzero elements in a vector $\vv$, that is $\|\vv\|_0 = \text{card}[\textmd{I}_{\vv}]$ - the {\it cardinality} of the {\it support} $\textmd{I}_{\vv}\triangleq\{l \in [L]:~\vv(l)\neq 0\}$ of a vector $\vv \in \mathbb R^{L\times 1}$. A signal $\vx$ is said to be $\kappa$-sparse in $\mQ$ if $\|\vs\|_0\leq \kappa$.

As is well known, the sparse recovery problem (\ref{f-1}) is NP-hard and has been investigated for many years. Determining how to solve problem (\ref{f-1}) {\it efficiently} and {\it accurately} is crucial for a wide range of sparsity-related problems. To the best of our knowledge, this remains an open problem, and there is still a significant need for more effective algorithms.

\subsection{Related work}
To accurately solve the sparse recovery problem (\ref{f-1}), the very first question to raise is: when does it have a unique solution $\vs^\star$? Let $\text{spark}[\mQ]$ denote the smallest number of columns of $\mQ$ that are linearly dependent.  It is well known that the following result~\cite{donoho2003optimally} plays a key role in sparse recovery: Let $\vx$ be any $\kappa$-sparse vector in $\mQ$, that is there exists a $\tilde \vs$  with $\|\tilde \vs\|_0\leq  \kappa$ such that $\vx=\mQ \tilde \vs$. Then, problem (\ref{f-1}) has one and only one solution, $\vs^\star = \tilde \vs$, if and only if $\mQ$ satisfies
\e  \text{spark}[\mQ]>2\kappa. \label{cs-2}\ee
Let $\kappa_{\mQ}$ be the largest integer that is not bigger than $(\text{spark}[\mQ]-1)/2$. Then, any $\kappa$-sparse solution $\tilde \vs$ of $\vx=\mQ\vs=\mQ\tilde \vs$ is the unique solution of (\ref{f-1}) as long as $\kappa \leq \kappa_{\mQ}$, which is assumed  in the sequel.

Another very important concept is the {\it restricted isometry property} (RIP), introduced in~\cite{candes2005decoding}, which is defined as follows:
a matrix $\mQ$ is said to satisfy the $(\kappa,\delta)$-RIP with $0\leq \delta <1$ if
\e (1-\delta)\|\vs\|^2_2\leq \|\mQ \vs\|^2_2\leq (1+\delta)\|\vs\|^2_2,~~~\forall~\vs\in \Sigma_{\kappa}, \label{RIP-1}\ee
where  $\Sigma_{\kappa}$ denotes the set of all $\kappa$-sparse signals. It can be seen  that if $\mQ$ satisfies the $(2\kappa, \delta)$-RIP, then any $2\kappa$ columns of $\mQ$ are linearly independent, and hence condition~(\ref{cs-2}) holds. Therefore, if $\vx = \mQ \vs^\star$ with $\|\vs^\star\|_0\leq \kappa$, then, the underdetermined system $\vx =\mQ\vs$ has a unique $\kappa$-sparse solution, which is $\vs^\star$. 

As an NP-hard problem, (\ref{f-1}) is not tractable due to the $l_0$  norm. To address this, a convex relaxation is commonly employed. In particular, two classes of methods are popularly used: 
\subsubsection{Iteratively re-weighted approaches}
This class includes basis pursuit (BP)-based algorithms, which rely on replacing the $l_0$ norm with $l_1$-minimization. This approach regularizes problem (\ref{f-1}) with the following problem~\cite{chen2001atomic,candes2005decoding}:
\begin{align}
\vs^\ast \triangleq \arg\min_{\vs}\|\vs\|_1~~~\text{s.t}~~~\vx = \mQ\vs.
\label{BP-1x}
\end{align}
This is a convex problem and there exist standard algorithms designed to solve such a problem efficiently, particularly when the problem's dimension is manageable. Denote $\text{BP}_C$ as the algorithm implemented using MATLAB command {\it linprog.m}.

Note that 
   $\|\vs\|_0=\sum^L_{l=1}\text{sgn}[|\vs(l)|]$,
 where $\text{sgn}[t]$ is the sign function. Thus, the $l_1$-minimization-based BP approach mentioned above is just an approximation of  $\|.\|_0$, where $\text{sgn}[|t|]$ is replaced with $\phi_1(t)\triangleq |t|$. 

To further improve the sparse recovery accuracy of the BP-based approach, a better approximating function $\phi(t)$ can be used, such as: 
$ \phi_2(t)\triangleq \text{log}(|t|+\epsilon); ~
\phi_3(t)\triangleq (|t|^2+\epsilon)^{q/2},~~0<q<1, 
$
where $\epsilon >0$ is a parameter used to regularize the singularity at $t=0$.
With this function, we define
$\|\vs\|_{\chi}\triangleq \sum^L_{l=1}\phi(\vs(l)).$
The corresponding sparse recovery problem is formulated as the following constrained minimization
\begin{align*} 
\vs^\ast\triangleq \arg\min_{\vs}\|\vs\|_{\chi}~~\text{s.t.}~~\vx = \mQ \vs. 
\end{align*}
This non-convex problem is then addressed with two classes of iteratively re-weighted algorithms~\cite{gorodnitsky1997sparse,rao1999affine,daubechies2010iteratively,huang2015two,candes2008enhancing,wang2010sparse,chartrand2008iteratively}.  
These algorithms are based on a first-order approximation approach and can be unified as follows:
\eqn \vs^{(k+1)} &\triangleq& \arg\min_{\vs}\|\mD^{(k)}_p\vs\|^p_p~~~\text{s.t}~~~\vx = \mQ\vs\nonumber, \\
 \mD^{(k+1)}_p &=&\text{diag}\{\vw_p[\vs^{(k+1)}]\},~~~~~~~~~p=1,2, \label{IR_BP-LS}\eeqn  
where $\vw_p[\vs]\in \mathbb R^{L\times 1}$ is determined by the weighting scheme. For $p=1$, the formulation \eqref{IR_BP-LS} is referred to as the iteratively re-weighted BP (IRBP)~\cite{candes2008enhancing,wang2010sparse,huang2015two}, and for $p=2$, it corresponds to the iteratively re-weighted least squares (IRLS)~\cite{daubechies2010iteratively,chartrand2008iteratively,gorodnitsky1997sparse,rao1999affine}.

In general, the weighting $\vw_p[\vs]$ is determined by  $\phi(t)$.  For example, the $l$th element of 
$\vw_1[\vs]$ is given by 
$$\vw_1[\vs(l)]=\frac{1}{|\vs(l)|+\epsilon}$$
in~\cite{candes2008enhancing}, where $\phi(t)=\phi_2(t)$  was used for IRBP, while with $\phi(t)=\phi_3(t)$ for IRLS
$$\vw_2[\vs(l)]=(|\vs(l)|^2+\epsilon)^{(q/2-1)/2}$$
   used in \cite{chartrand2008iteratively}. 

The IRBP-based algorithms provide a powerful tool for solving sparse recovery problems. However, these algorithms are computationally demanding, especially for large-scale scenarios. To reduce computational complexity, an $l_1$-$l_2$ minimization-based reformulation of \eqref{BP-1x} was proposed in  \cite{chen1998atomic,tibshirani1996regression}, and a class of iterative algorithms, including FISTA, was derived \cite{beck2009fast,mousavi2017learning}. 

The IRLS method is popularly used for non-smooth cost functions by solving a sequence of quadratic problems as specified in \eqref{IR_BP-LS} with $p=2$. The solution to the 1st equation of  \eqref{IR_BP-LS} is given by
\begin{align}
\vs^{(k+1)}=(\mD^{(k)}_2)^{-2}\mQ^\top[\mQ(\mD^{(k)}_2)^{-2}\mQ^\top]^{-1}\vx.\label{matrix-inv-LS-cla}    
\end{align} 
Note that the algorithms \eqref{IR_BP-LS}  are $\epsilon$-sensitive, and a time-varying $\epsilon$ is crucial for these algorithms to achieve {\it state-of-the-art} recovery performance.  In \cite{chartrand2008iteratively}, an $\epsilon$-regularization strategy was proposed, where   \eqref{IR_BP-LS} with $p=2$ is iteratively run with a monotonically decreasing sequence $\{\epsilon_j: 10^{-8}\leq \epsilon_j\leq 1\}$. In this iterative procedure, $\vs_j$ - the solution given by \eqref{IR_BP-LS} with $\epsilon_j$, is used as the initial iterate for the next iteration. We denote such an iterative procedure as $\text{IRLS}_{C}$. 


\subsubsection{Greedy methods}
By nature, sparse recovery aims to identify the support $\textmd{I}_{\vs^\star}$ of the $\kappa$-sparse vector $\vs^\star$ underlying the measurement $\vx$. Thus, one sparse algorithm differs from another in how it detects the support. The iteratively re-weighted methods estimate the support $\textmd{I}_{\vs^\star}$ through a global search in the $\vs$-space and then select the indices of the $\kappa$ largest entries in magnitude from their minimizer $\vs^\ast$.

The greedy algorithms, as described in~\cite{mallat1993matching,tropp2004greed}, intends to address the sparse recovery problem~(\ref{f-1}) by solving
the following alternative problem:
\eqn \vs^{\#} \triangleq \min_{\vs}\|\vx -\mQ \vs\|^2_2~~~\text{s.t.}~~~\|\vs\|_0\leq \kappa,\nonumber\eeqn
where $\kappa$ is the given sparsity level.  The $\kappa$ non-zero entries of $\vs^{\#}$ are identified sequentially through a procedure that minimizes residual.
Such a method is known as the orthogonal matching pursuit (OMP) method.

The OMP algorithm extracts the $\kappa$ indices one by one iteratively. The $k$th atom, indexed by $i_k$, is identified as follows:
\e i_k\triangleq \arg\min_{i,\beta}\|\vr_{k-1}-\beta \mQ(:,i)\|^2_2. \label{insert-1}\ee
Here, $\vr_{k-1}$ is the residual obtained at the $(k-1)$th iteration:\footnote{Throughout this paper, we use MATLAB notations:  $\vv(i)$ denotes the $i$th entry of a vector $\vv$, while $\mA(i,j),~  \mA(:,j)$, and $\mA(i,:)$ denote the $(i,j)$th entry, the $j$th column, and the $i$th row of a matrix $\mA$, respectively. Furthermore, let $\mathcal I$ be a subset of $[L]=\{1,2,\cdots,L\}$. $\vv(\mathcal I)$ is defined as the sub-vector of the vector $\vv$ with dimension $L$, obtained by excluding those entries whose indices do not belong to $\mathcal I$.}
$\vr_{k-1}=\vx - \mQ(:,\mathcal I_{k-1})\vbeta_{k-1},$
where $\mathcal I_{k-1}$ is the set of the $k-1$ indices detected before the $k$th iteration and $\vbeta_{k-1}$ is the associated coefficient vector obtained in the $(k-1)$th iteration.


\subsection{Problem formulation and contribution}\label{sec-1-2}
It should be pointed out that 
the iteratively re-weighted  algorithms provide {\it state-of-the-art} sparse recovery performance.  However, they are computationally demanding. In particular, IRLS-based algorithms, though much faster than the IRBP-based ones, can encounter numerical issues when the $\mQ$ matrix is ill-conditioned (See \eqref{matrix-inv-LS-cla}). By contrast, OMP is significantly more efficient.
The most computationally demanding stage involves solving the least squares problem to update the non-zero entries. A class of iterative methods, such as gradient pursuit (GP) and conjugate gradient pursuit (CGP), can be used to address this issue, as described in~\cite{blumensath2008gradient}. To improve the recovery accuracy and update the index set of the $\kappa$ non-zero entries more efficiently, one approach is to select multiple entries at once, rather than just one as in OMP. This leads to the development of stage-wise OMP (StOMP)~\cite{donoho2012sparse}, compressive sampling match pursuit (CoSaMP)~\cite{needell2009cosamp}, and stagewise weak GP (StWGP)~\cite{blumensath2009stagewise}. A comprehensive survey of these issues as well as performance guarantees can be found in~\cite{blumensath20128}.

\textcolor{black}{As is well-known, the greedy methods can not yield a comparable sparse recovery performance to the iteratively re-weighted ones. Suppose that $\mathcal I_{k-1}$ has its $k-1$ indices that all fall within the support $\textmd{I}_{\vs^\star}$. Then, $\vr_{k-1}= \mQ\vs - \mQ(:, \mathcal I_{k-1})\vbeta_{k-1}\triangleq \mQ\tilde \vs$, and this new system has a unique sparsest
solution, denoted as $\tilde \vs^\star$, with $\|\tilde \vs^\star\|_0 \leq \kappa-(k-1)$. The $l_2$-norm based minimization (\ref{insert-1}) can not ensure that the obtained index $i_k$ falls into $\textmd{I}_{\vs^\star}$ because the cost function is a complicated function of the true indices to be detected and such a function changes with $\vbeta_{k-1}$. This is the main reason why the OMP algorithm cannot generally yield a sparse recovery as accurate as the iteratively re-weighted algorithms, prompting the proposal of its variants such as CoSaMP and StOMP.}

\textcolor{black}{The question we ask ourselves: Can we transform the system $\vr_{k-1}=\mQ\tilde \vs$ into $\vs$ domain by characterizing its set of solutions and then work directly on the sparse recovery problem? If an estimate of $\tilde \vs^\star$ is achieved, the next index $i_k$ can be selected as the one that corresponds to the largest entry in magnitude of this estimate. Intuitively,  by doing so the accuracy of index detection can be improved. This motivates us to deal with the sparse recovery problem on the $\vs$-space, 
instead of the measurement $\vx$-space.}
\SL{A more detailed discussion of this idea is provided in Section~\ref{sec:pro_greedy}.
}

The main objective of this paper is to derive a class of greedy algorithms that operate directly in the $\vs$-domain and outperform classical greedy algorithms (i.e., OMP, CoSaMP) in terms of sparse recovery accuracy while maintaining similar implementation complexity. Additionally, these algorithms offer greater robustness against the numerical instability caused by $\mQ$ and interferences in the signal $\vx$ compared to the classical algorithms, especially the IRLS-based ones. 
Our contributions are summarized as follows.

\begin{itemize}\item Based on a characterization of solutions to equation~(\ref{Synthetic-model}), we propose a class of greedy sparse recovery approaches for a given sparsity measure. Such an approach allows us to work directly in the $\vs$-space;
\item Two algorithms, denoted as $\text{Alg}_{GL2}$ 
and $\text{Alg}_{GL1}$, 
are derived based on $l_2$ and $l_1$, respectively. 
Despite having nearly the same computational complexity, $\text{Alg}_{GL2}$ significantly outperforms the classical OMP algorithm and is comparable to the classical BP algorithm in terms of recovery accuracy. Meanwhile, the $\text{Alg}_{GL1}$ algorithm significantly surpasses the classical OMP, BP$_C$, and CoSaMP algorithms in terms of sparse recovery accuracy;
\item An improved version of $\text{Alg}_{GL1}$ has been derived, denoted as the $\text{Alg}_{GLQ}$ algorithm, in which $l_q$-minimization is used to enhance the sparsity. Fast variants of the algorithms $\text{Alg}_{GL1}$ and  $\text{Alg}_{GLQ}$, denoted as $\text{Alg}^F_{GL1}$ and $\text{Alg}^F_{GLQ}$, are developed to accelerate $\text{Alg}_{GL1}$ and $\text{Alg}_{GLQ}$ by selecting multiple atoms per iteration;
\item It is shown that the proposed algorithms such as $\text{Alg}_{GL1}$, $\text{Alg}_{GLQ}$ and their fast variants can be used to deal with low-rank interferences in the measurements effectively, which owes to the proposed characterization and is a significant advantage over the classical algorithms. \end{itemize} 
All claims have been validated through extensive numerical simulations on both synthetic and image data. 


The paper is organized as follows. In Section~\ref{sec-2}, we derive a characterization of the solution set for the linear system $\vx =\mQ \vs$. Based on this characterization, we propose a greedy approach to sparse recovery. Two specified algorithms, denoted as $\text{Alg}_{GL2}$ and $\text{Alg}_{GL1}$, are presented in this section. Section~\ref{sec-3} analyzes the performance of the proposed algorithms.  An improved version of $\text{Alg}_{GL1}$, denoted as $\text{Alg}_{GLQ}$, is derived in Section~\ref{sec-4}, where the $l_q$-minimization is employed to enhance the sparsity. Fast variants of both $\text{Alg}_{GLQ}$ and $\text{Alg}_{GL1}$ are also derived in this section. Numerical examples and simulations are presented in Section~\ref{sec-5} to validate the theoretical results, evaluate the performance of the proposed methods, and compare them with some existing approaches. To end this paper, some concluding remarks are given in Section~\ref{sec-6}.


\section{The Proposed Approach and Algorithms}\label{sec-2}
Our proposed greedy approach is based on the following characterization of the solutions to the system~\eqref{Synthetic-model}.

Let $\mQ=\mU\MAT{cc}\mSigma& {\mzero}\\ {\mzero}& {\mzero}\mat \mV^\top$ be a singular value decomposition (SVD) of $\mQ$, where the diagonal matrix $\mSigma \succ 0$ has dimension $ \tilde N \times \tilde N$ with $\tilde N \leq N$. It then follows from $\mQ\vs = \vx$ that
$$\MAT{cc}\mSigma& {\mzero}\\
{\mzero}& {\mzero}\mat \mV^\top\vs = \mU^\top\vx.$$
Denote $\mV_1\triangleq \mV(:,1:\tilde N)$, $\mW\triangleq \mV(:,\tilde N+1:L)$, and $\mU_1\triangleq \mU(:,1:\tilde N)$. With $\mV^\top\vs = \MAT{c}\bar{\vs}_1\\ \vz\mat$, we can show that any solution $\vs$ to (\ref{Synthetic-model}) is given by $\vs = \mV \MAT{c} \bar{\vs}_1\\ \vz\mat $ and hence
is of the form $\vs=\mV_1\mSigma^{-1}\mU^\top_1\vx+ \mW\vz$, i.e.,
\eqn
\vs
\triangleq \mV_1\vx_c + \mW \vz\triangleq \vs_0(\vx) +\mW\vz\triangleq \vs(\vz,\vx),\label{characterization}\eeqn
 where $\vx_c=\mSigma^{-1}\mU^\top_1\vx\triangleq \mT_0\vx$. Consequently, the solution set $\mathcal S_0 \triangleq \{\vs(\vz,\vx)\}$ of 
system \eqref{Synthetic-model} is completely characterized by the vector variable $\vz \in \mathbb R^{(L-\tilde N)\times 1}$ for a given $\vx$. The sparse recovery problem \eqref{f-1} can then be converted into the following equivalent form\footnote{For simplicity, $\vs_0(\vx)$ and $\vs(\vz,\vx)$ are sometimes denoted as $\vs_0$ and $\vs(\vz)$, respectively, in the sequel.}
\eqn \vz^\star &\triangleq& \arg\min_{\vz}\|\vs_0+\mW\vz\|_0\quad\mapsto ~~~~\vs^\star = \vs_0+\mW\vz^\star.\label{restart-1} \eeqn
The alternative formulation transforms the classical problem~(\ref{f-1}) from a constrained minimization to an unconstrained one, allowing us to directly address sparse recovery in the $\vs$-space of dimension $L\times 1$.

A traditional sparse recovery algorithm aims to find the $\kappa$-sparse solution $\vs^\star$ for each measurement $\vx$, where the system is parametrized by $\mQ$ according to the model (\ref{Synthetic-model}). In this paper, we propose a class of greedy algorithms that obtain the solution based on the proposed model (\ref{characterization}), where the system is parametrized by $(\mW, \mV_1)$. The input $\vs_0$ is computed as $\vs_0=\mV_1 \vx_c=\mV_1\mT_0\vx$, serving as an initial estimate of $\vs^\star$.
The parameters $(\mW, \mV_1, \mT_0)$, which are equivalent to $\mQ$, are determined at the design stage. When deriving the algorithms, we assume that 
the triple $(\mW, \mV_1, \vs_0)$
is available, similar to $(\mQ,\vx)$ in traditional methods. 

\subsection{The proposed greedy approach}\label{sec:pro_greedy}
 Suppose that $i_1$ is a detected index, expected to be one of the elements in $\textmd{I}_{\vs^\star}$. Let's consider how to detect the next index, say $i_2$, in a greedy manner.

First of all, it follows from $\vx =\mQ \vs$ that the residual is given by
$\vx - \beta \mQ(:,i_1)=\mQ \vs - \beta\mQ(:,i_1)$
or more simply $\vx - \beta \mQ(:,i_1)=\mQ \vs_1$,
where $\vs_1$, according to the proposed characterization~(\ref{characterization}), can be expressed as
\begin{align*}
\vs_1 &= \mV_1\mT_0(\vx - \beta \mQ(:,i_1)) +\mW\vz =\vs_0 +\mW\vz  - \beta \mV_1\mT_0\mQ(:,i_1) =\vs(\vz) - \beta \mV_1\mT_0\mQ(:,i_1), 
\end{align*}
where $\vs(\vz)=\vs_0 +\mW\vz \in \mathcal S_0$, as defined in  (\ref{characterization}).

Recall that $\mQ=\mU_1\mSigma\mV_1^\top$. Together with $\mT_0=\mSigma^{-1}\mU^\top_1$ and $\vs_0=\mV_1\mT_0\vx=\mV_1\mT_0\mQ\vs^\star$, we obtain
\eqn \vs_0 =\mV_1\mV^\top_1\vs^\star\triangleq  \bar \mQ\vs^\star,\quad
\mV_1\mT_0\mQ(:,i_1)=\bar \mQ(:,i_1),\label{HD-system-0}\eeqn
where $\bar \mQ \triangleq \mV_1\mV^\top_1$.
Define a set $\mathcal S_1 \triangleq \{\vs_1: \vs_1=\vs(\vz)  - \beta \bar \mQ(:,i_1)\}$ that is characterized by $(\vz, \beta)$. 

\SL{It is important to note that since we assume $\mathcal S_0$ has a unique $\kappa$-sparse vector $\vs^\star = \vs_0 + \mW\vz^\star$, $\mathcal S_1$ possesses the following properties: 
\begin{itemize}\item having a unique $(\kappa-1)$-sparse element $\vs^\star_1$, if and only if $i_1\in \textmd{I}_{\vs^\star}$ and $\beta=\vs^\star(i_1)\triangleq \beta^\star_1$;
\item $\|\vs_1\|_0\geq \kappa, ~\forall~\vs_1\in \mathcal S_1$, if $i_1 \notin \textmd{I}_{\vs^\star}$.
\end{itemize}
Thus, any solution of
\e \{\tilde i, \tilde \beta,\tilde \vz\}\triangleq \arg\min_{i,\beta,\vz}\|\vs(\vz)-\beta\bar\mQ(:,i)\|_0 \label{Extr-1}\ee
satisfies $\tilde i \in \textmd{I}_{\vs^\star}, 
\tilde \beta =\vs^\star(\tilde i)$, and $\tilde \vz = \vz^\star$.}

Assume that $i_1 \in \textmd{I}_{\vs^\star}$.  We note that problem~(\ref{Extr-1}) with $i=i_1$ results in $\tilde \vz=\vz^\star$ and $\tilde \beta = \beta^\star_1$. Then, the corresponding  $\vs^\star_1= \vs^\star - \vs^\star(i_1)\bar\mQ(:,i_1)$
is the unique $(\kappa-1)$-sparse and the sparsest element in set $\mathcal S_1$. This fact implies that to detect the next index in $\textmd{I}_{\vs^\star}$, say $i_2$, we can first identify the sparsest element in $\mathcal S_1$. Then, $i_2$ can be determined as the index corresponding to \SL{any non-zero entry, say} the largest entry in magnitude, of $\vs^\star_1$.
\SL{Therefore, the $l_0$-based index selection strategy in~\eqref{Extr-1}, built on our proposed characterization~\eqref{characterization}, yields a more reliable identification of the support than the formulation~\eqref{insert-1} used in the classical greedy methods.}

It is based on this observation that we propose the following greedy method, whose $(k+1)$th iteration  involves

\eqn \left\{\begin{array}{rcl} \{\tilde \vz_k, \tilde\vbeta_k\} &\triangleq& \arg\min_{\vz,\vbeta}\|\vs(\vz)   - \bar \mQ_{k}\vbeta\|_{\chi} \quad
\mapsto~\tilde \vs_{k} = \vs(\tilde \vz_{k})- \bar \mQ_{k}\tilde\vbeta_{k},\\
i_{k+1} &\triangleq& \arg\max_{\forall ~i_l\in \mathcal I^c_{k}}|\tilde \vs_{k}(i_l)|
\quad~~~ \quad\mapsto~~\mathcal I_{k+1}=\mathcal I_k\cup i_{k+1}, \end{array}\right.\label{Greedy-Approach}\eeqn
where  $\bar \mQ_{k}\triangleq \bar \mQ(:,\mathcal I_{k})$ with $\mathcal I_{k}\triangleq \{i_1, i_2, \cdots,i_k\}$ being the set of indices detected before, and $\mathcal I^c_{k}$ its complement.\footnote{Throughout this paper, $\mathcal I_k$ denotes a set of indices when $k$ is an integer, while, as defined before, $\textmd{I}_{\vv}$ denotes the support of a vector $\vv$.} Here,  $\|.\|_{\chi}$ denotes the $l_0$ norm or one of its  approximating measures.

Note that the pair $(\tilde \vs_{k}, \tilde\vbeta_{k})$ in (\ref{Greedy-Approach}) corresponds to an $\vs_k$ that satisfies $\vx = \mQ\vs_k$, where
\e \vs_k = \tilde \vs_{k} + \tilde\vbeta^\ast_{k},\label{Insert-X}\ee
and $\tilde\vbeta^\ast_{k}$ is $k$-sparse with $\tilde\vbeta^\ast_{k}(\mathcal I_{k})=\tilde\vbeta_{k}$.

Running (\ref{Greedy-Approach}) for $\kappa$ iterations, 
the $\kappa$-sparse estimate of $\vs^\star$, denoted as $\hat \vs^\star$,  is given by
$ \hat \vs^\star(\mathcal I_{\kappa}) = \tilde \vbeta_{\kappa}$.  
It will be shown in the next section that if $\tilde \vs_{\kappa}=\vzero$, then
$\mathcal I_{\kappa} =\textmd{I}_{\vs^\star}$, and $\tilde \vbeta_{\kappa} = \vs^\star(\mathcal I_{\kappa}),$
which implies that the unique $\kappa$-sparse solution $\vs^\star$ of $\vx = \mQ\vs$ has been obtained.

Although the  $\|.\|_0$ norm is the best measure to deal with sparsity-related problems,  one of its approximating measures has to be used, as the $\|.\|_0$ norm is NP-hard and hence not tractable. In the next two subsections, we will further elaborate on the proposed greedy approach (\ref{Greedy-Approach}) with two approximating
measures.

\subsection{An $l_2$-relaxation}
Now, consider $\chi =2$, that is, the $l_2$-relaxation of $l_0$. Denote $\vs_{k} = \vs(\vz)  - \bar \mQ(:,\mathcal I_{k})\vbeta$. We have
$$\|\vs_{k}\|^2_2 = \|\vs_0  - \bar \mQ(:,\mathcal I_{k})\vbeta\|^2_2+\|\mW\vz\|^2_2$$
due to the fact that $\vs_0=\bar \mQ\vs^\star$, where $\bar \mQ=\mV_1\mV^\top_1$, as defined in (\ref{HD-system-0}),  and   $\mV_1^\top\mW=\mzero$.
Then, the proposed greedy approach (\ref{Greedy-Approach}) turns to
\e \left\{\begin{array}{rcl}\tilde \vbeta_{k} &\triangleq&\arg\min_{\vbeta}\| \vs_0   - \bar \mQ(:,\mathcal I_{k})\vbeta\|^2_2\quad\mapsto~~\tilde \vs_{k} = \vs_0 - \bar \mQ(:,\mathcal I_{k})\tilde \vbeta_{k}\\
i_{k+1} &\triangleq& \arg\max_{\forall ~i_l\in \mathcal I^c_{k}}|\tilde \vs_{k}(i_l)|\quad \quad\quad~  \mapsto~~\mathcal I_{k+1}=\mathcal I_k\cup i_{k+1}\end{array}\right. \label{Greedy-Approach-OMP-HD}\ee
as $\tilde  \vz_{k}=\vzero,~\forall~k$.
We name the above approach $\text{Alg}_{GL2}$, where the subscript ``$_{GL2}$" stands for greedy method with indices selected using $l_2$ minimization. We briefly denote the approach as  $\hat \vs^\star = \text{Alg}_{GL2}(\vs_0,\mV_1,\kappa)$, where $\vs_0=\mV_1\mT_0\vx$, as mentioned before, is the initial estimate of $\vs^\star$.

We realize by noting  $\vs_0=\bar \mQ \vs^\star$ that the proposed $\text{Alg}_{GL2}$ algorithm~(\ref{Greedy-Approach-OMP-HD}) is closely related to the classical OMP algorithm applied to the following higher dimensional system (than $\vx = \mQ\vs$)
\e \vs_0 = \bar \mQ \vs.  \label{low-to-high}\ee
It can be seen that the $\text{Alg}_{GL2}$ algorithm differs from the classical OMP applied to (\ref{low-to-high}) in how it detects the next index $i_{k+1}$.


\begin{Remark}
\label{Remark 2.2}
It should be pointed out that, although the
transformed system $\vs_0 = \bar \mQ\vs$ is theoretically equivalent to the original $\vx =\mQ\vs$, the proposed $\text{Alg}_{GL2}$ algorithm significantly outperform the classical $\text{OMP}_C(\vx,\mQ,\kappa)$ algorithm - the $\text{OMP}$ algorithm applied directly to $\vx =\mQ\vs$. The most important reason for this is the ``residual" in the transformed system being $\tilde \vs_k$, instead of the  $\vr_k$ in the classical greedy approach. More details will be analyzed in the next section.
\end{Remark}

\subsection{An $l_1$-relaxation}
Note that $\vs_{k} = \vs_0 +\mW \vz  - \bar \mQ(:,\mathcal I_{k})\vbeta$ can be rewritten as $\vs_{k} = \vs_0 + \mOmega_k\bar \vz$, where
\eqn \mOmega_k&\triangleq &\MAT{cc}\mW&-\bar\mQ(:,\mathcal I_{k})\mat, ~\bar \vz \triangleq \MAT{c}\vz\\ \vbeta\mat.\nonumber\eeqn
With $\chi =1$, the proposed greedy approach (\ref{Greedy-Approach}) becomes
\e \left\{\begin{array}{rcl}\bar \vz_{k} &\triangleq& \arg\min_{\bar \vz}\|\vs_0 +\mOmega_k\bar \vz\|_1\quad \mapsto~~ \tilde \vs_{k} = \vs_0 +\mOmega_k \bar \vz_{k},\\
 i_{k+1} &\triangleq& \arg\max_{\forall ~i_l\in \mathcal I^c_{k}}|\tilde \vs_{k}(i_l)|\quad \mapsto~~\mathcal I_{k+1}=\mathcal I_k\cup i_{k+1},\end{array}\right. \label{Greedy-Approach-BP}\ee
which is a greedy $l_1$-based algorithm for sparse recovery.

As seen, such an iterative algorithm involves solving a series of $l_1$-minimizations of the form
\e \tilde \vz \triangleq \arg\min_{\vz}\|\vs_0+\mOmega\vz\|_1. \label{alternative-2x}\ee   
As shown in \cite{candes2005decoding}, the above problem can be converted as a {\it linear programming} problem and hence can be solved with one of the standard convex optimization techniques, such as the MATLAB command $linprog.m$.
%
For convenience, we denote the corresponding algorithm as $\text{BP}_{alt}$:
 $[\tilde \vz, \tilde \vs] = \text{BP}_{alt}(\vs_0, \mOmega)$ with the residual $\tilde \vs = \vs_0+\mOmega \tilde \vz$.
%

The outline of the proposed greedy $l_1$-based algorithm~\eqref{Greedy-Approach-BP}, denoted as $\text{Alg}_{GL1}$, is then 
presented in Algorithm~\ref{alg:GL1}. For convenience, we denote this algorithm as
$\hat \vs^\star = \text{Alg}_{GL1}(\vs_0, \mW, \mV_1,\kappa).$


Before proceeding to the next section, we first evaluate the performance of the proposed greedy algorithms through the following numerical experiment. We generate a sequence $\{\vx_j\}$ of $J=1,000$ samples with $\vx_j = \mQ\vs_j, ~\forall~j$. Here,
$\mQ^{64\times 128}$ - a (column) $l_2$-normalized matrix, and $\{\vs_j\}$ with $\|\vs_j\|_0=20$ are generated randomly. We run each of the four algorithms:
$\text{OMP}_C(\vx,\mQ,\kappa)$, $\text{BP}_C$ - the classical BP method, 
$\text{Alg}_{GL2}(\vs_0,\mV_1,\kappa)$
and $\text{Alg}_{GL1}(\vs_0, \mW, \mV_1,\kappa)$, where $\vs_0 = \mV_1\mT_0\vx$ with $\vx=\vx_j$,
for all the 1,000 samples $\{\vx_j\}$.

\begin{algorithm}[H]
\caption{A greedy $l_1$-based method for sparse recovery ($\text{Alg}_{GL1}$)}\label{alg:GL1}
\begin{algorithmic}
\State\textbf{Inputs:} $(\vs_0, \mW,\mV_1)$ and $\kappa$.
\State\textbf{Initials:} 
$\mathcal I=[~]$. Run
$[\tilde \vz,\tilde \vs]=\text{BP}_{alt}(\vs_0, \mW)$,
and assign the residual $\tilde \vs$ to $\vs$;\footnotemark
\For {$k=1:\kappa$}
\State - Checking whether $\vs$ is the desired solution: Let $|\vs|_{\kappa+1}$ denote the $(\kappa+1)$-th entry of the  re-arranged sequence of $\{|\vs(l)|\}$ sorted in a descending order.~~
${\bf If} ~~|\vs|_{\kappa+1} \leq 10^{-4}, ~ {\bf break}~~ {\it for-loop}.$
\State - Updating index $\mathcal I$: Let $i_m$ be the index such that $|\tilde \vs(i_m)|=\max_{i}\{|\tilde \vs(i)|\}$. Then,
   $\mathcal I ~\leftarrow~ [\mathcal I~~i_m]$.
\State - Updating the residual $\tilde \vs$ and $\tilde \vbeta$: Compute $\mOmega = \MAT{cc}\mW& -\mV_1(\mV_1(\mathcal I,:))^\top\mat$, then run    $[\vz,\tilde \vs] = \text{BP}_{alt}(\vs_0,\mOmega),$
set $\tilde \vbeta = \vzero$ and $\tilde \vbeta(\mathcal I)$ with the last $\text{length}(\mathcal I)$ entries of $\vz$; 
\State - Updating $\vs$ using  (\ref{Insert-X}) with the pair $(\tilde \vs, \tilde \vbeta)$:~~
$\vs = \tilde \vs + \tilde \vbeta.$
\EndFor
\State\textbf{Output:} $\hat \vs^\star=\vs$.
\end{algorithmic}
\end{algorithm}

\footnotetext{\SL{Here, $\tilde \vs$ is used as an initial estimate of $\vs^\star$ and can be obtained using algorithms other than $\text{BP}_{alt}$.}}

\SL{Let $\textmd{I}_{\lambda}[\vv]$ denote the index set of the $\lambda$ largest absolute entries of $\vv$. 
Let $\hat \vs^\star$ be an estimate of the true signal $\vs^\star$, both being $\kappa$-sparse. 
A recovery is considered successful if the estimate $\hat \vs^\star$ obtained by an algorithm satisfies
$ \textmd{I}_{\kappa}[\hat \vs^\star]=\textmd{I}_{\vs^\star}.$
We define the signal-to-noise ratio of such an estimation as
$\varrho_{snr}\triangleq 10\times \text{log}_{10}\frac{\|\mQ\vs^\star\|^2_2}{\|\mQ(\vs^\star-\hat \vs^\star)\|^2_2}~~~(\text{dB})$. 
When some of the $\kappa$ nonzero entries of $\vs^\star$ are very small, these entries have a little contribution to $\vx = \mQ\vs^\star$ but it is very difficult to correctly identify the indices of these entries. 
Therefore, in this paper, we consider a recovery to be successful if either all the indices of $\vs^\star$ are detected or the corresponding $\varrho_{snr}$ exceeds 60 dB.}\footnote{
\SL{Note that the error $\|\hat{\vs}^\star - \vs^\star\|_2$ is not a reliable indicator of successful sparse recovery. Even when this error is large, a perfect recovery can still be achieved as long as the support is correctly identified. In this case, the exact coefficients of $\vs^\star$ can be recovered by solving a least-squares problem on the identified support.}
}

\begin{wraptable}{r}{0.5\textwidth}  
\vspace{-1.0cm}
\centering
\caption{Recovery rates for the four algorithms. }
\label{tab:recovery_rates_horizontal}
\begin{tabular}{|c|c|c|c|c|}
\hline
& $\text{OMP}_C$ & $\text{BP}_C$ & $\text{Alg}_{GL2}$& $\text{Alg}_{GL1}$ \\ \hline
$\varrho_{ok}$ & 65.80\% & 94.30\% & 97.40\% & 100\% \\ \hline
\end{tabular}
\vspace{-0.3cm}
\end{wraptable}


The rate of successful recovery for an algorithm running $J$ samples is defined as
$\varrho_{ok}\triangleq J_{ok}/J$,
where $J_{ok}$ is the number of successful recoveries out of the $J$ samples.
With the above setting, the successful recovery rates for the four algorithms are given in Table~\ref{tab:recovery_rates_horizontal}.




\section{Performance Analysis of the Proposed Greedy Approach}\label{sec-3}
From the numerical example presented above, we observe that the proposed $\text{Alg}_{GL2}$ algorithm significantly outperforms the classical $\text{OMP}_C$ and even surpasses $\text{BP}_C$. Meanwhile, as expected, the proposed $\text{Alg}_{GL1}$ achieves a perfect sparse recovery across all 1,000 samples. These results demonstrate that our proposed greedy approach successfully achieves the objective raised in Section~\ref{sec-1-2}. In this section, we conduct a detailed analysis of the proposed approach and the resulting algorithms.

\subsection{Approach analysis}
As understood, greedy methods, including our proposed approach (\ref{Greedy-Approach}), are designed based on minimizing a residual. Consequently, the generated sequence of residuals is non-increasing.
Although the residuals converge, most greedy algorithms are expected to yield the unique solution $\vs^\star$ within a finite number of iterations.

First of all, we present the following result associated with the proposed greedy approach (\ref{Greedy-Approach}).

\begin{theorem}\label{theorem-Greedy-Alt}
Let $\mathcal I_k$ denote the set of indices detected, and let $(\tilde \vbeta_k, \tilde \vz_k)$ be obtained using the proposed greedy approach (\ref{Greedy-Approach}). Furthermore, let $\tilde \vs_k = \vs(\tilde \vz_k)-\bar\mQ_k \tilde \vbeta_k$ be the residual, and denote $\hat \vs_k \in \mathbb R^{L\times 1}$ as the vector satisfying
$$\hat \vs_k(\mathcal I^c_k)=\vzero,~~\hat \vs_k(\mathcal I_k)=\tilde \vbeta_k.$$
Then, when $\tilde \vs_{k}=\vzero$ for $k \leq 2\kappa$, we have
$\hat \vs_k = \vs^\star$
as long as $\textmd{I}_{\vs^\star} \subseteq \mathcal I_k$.
\end{theorem}

\noindent{\bf Proof}: First of all, we note that $\tilde \vs_{k}=\vzero$, that is
$ \vs_0 +\mW \tilde\vz_k - \bar \mQ(:, \mathcal I_k)\tilde\vbeta_k=\vzero,$
which is equivalent to $\bar \mQ(\vs^\star -\hat \vs_k) + \mW \tilde\vz_k=\vzero$. It then follows from $\mW^\top\mV_1=\mzero$ and $\bar \mQ =\mV_1\mV^\top_1$ that
$   \tilde\vz_k=\vzero ~\text{and}~ \mV^\top_1(\vs^\star  - \hat \vs_k)=\vzero.$

We note that $\vs^\star-\hat \vs_k$ is $2\kappa$-sparse when the condition given in the theorem holds and $\text{spark}[\mV^\top_1]=\text{spark}[\mQ]>2\kappa$ due to the assumption of uniqueness on $\vs^\star$. Thus, $\mV^\top_1(:,\mathcal I_k)$ is of full (column) rank, and hence the above equation leads to $\vs^\star-\hat \vs_k=\vzero$. This completes the proof.
\cqfd

In fact, the claims in Theorem~\ref{theorem-Greedy-Alt} also hold for the classical $\text{OMP}_C$ algorithm, in which the upper bound for the number of iterations is set to $\kappa$. This means that $\text{OMP}_C$ must select an index that falls within $\textmd{I}_{\vs^\star}$, which precludes the possibility of $\text{OMP}_C$ making some mistakes in selecting indices. However, these mistakes can be rectified by taking more than $\kappa$ iterations~\cite{blumensath20128}. In general, for $\mathcal I_k$ with $\kappa < k \leq 2\kappa$, nothing can guarantee whether $\textmd{I}_{\vs^\star} \subseteq \mathcal I_k$ holds unless $\|\tilde \vbeta_k\|_0=\kappa$ is verified.

The primary advantage of the proposed greedy approach over classical methods, such as $\text{OMP}_C$ and $\text{CoSaMP}$,  lies in its index selection strategy. In the proposed greedy approach (\ref{Greedy-Approach}), the next index $i_{k+1}$ is chosen as the one corresponding to the largest absolute entry of the residual $\tilde \vs_k$. As demonstrated in 
Section~\ref{sec:pro_greedy},
this selection ensures that $i_{k+1}$ belongs to $\textmd{I}_{\vs^\star}$.
In contrast, $\text{OMP}_C$ determines $i_{k+1}$ by minimizing $\|\vr_k-\beta\mQ(:,i)\|^2_2$ with respect to $\beta$ for $i\in \mathcal I^c_k$. However, such a selection does not necessarily guarantee that $i_{k+1}$ belongs to $\textmd{I}_{\vs^\star}$, even if $\mathcal I_k \subseteq \textmd{I}_{\vs^\star}$, as the minimizer of $\|\vr_k-\beta\mQ(:,i)\|^2_2$ may not correspond to an index within $\textmd{I}_{\vs^\star}$.

Furthermore, let $\mathcal I_k$ be the detected indices and $\mathcal I^\ast_{k}$ be a subset of $\mathcal I_k$, which contains $k^\star$ elements with $\mathcal I^\ast_{k} \subseteq \textmd{I}_{\vs^\star}$. Using similar arguments to those in 
Section~\ref{sec:pro_greedy}, we have
\e \{\tilde \vbeta_k, \tilde \vz_k\}\triangleq \arg\min_{\vbeta,\vz}\|\vs(\vz)-\bar \mQ(:,\mathcal I_k)\vbeta\|_0,\nonumber\ee
for which the residual is
$$\tilde  \vs_k =\vs(\tilde \vz_k)-\bar \mQ(:,\mathcal I_k)\tilde \vbeta_k = \vs^{\star}_{k^\star}$$
 as long as
$k^\star > k-k^\star~~\Rightarrow~~k^\star > k/2$
holds, where $\vs^{\star}_{k^\star}(l)=0, ~l\in \mathcal I^\ast_{k}$ and otherwise, $\vs^{\star}_{k^\star}(l)=\vs^\star(l)$. This implies that with $\chi$ taking $l_0$, the index $i_{k+1}$ of the absolutely largest entry of the residual $\vs^\star_k$ should fall within $\textmd{I}_{\vs^\star}$, provided that more than half of the $k$ detected indices belong to  $\textmd{I}_{\vs^\star}$. In other words, when the first few indices are detected correctly, the proposed greedy approach tolerates a small number of mistakes in index detection. This is indeed a remarkable property. 

\begin{Remark}
\label{Remark 3.1} 
The conclusion drawn above is based on the use of the ideal sparsity measure $l_0$. Intuitively, the same conclusion is likely to hold with a high probability when a good approximating measure of $l_0$ is used. Indeed, the excellent performance of $\text{Alg}_{GL1}$, in which $l_1$ is used, appears to support this argument. This highlighted property suggests that algorithms yielding better performance than $\text{Alg}_{GL1}$ could be developed by using the proposed greedy approach with more suitable approximating measures $\chi$ of $l_0$. This will be elaborated further in the next section.
\end{Remark}

\subsection{Comparison between $\text{OMP}_C$ and $\text{Alg}_{GL2}$}
As argued above, the key to $\text{Alg}_{GL2}$ outperforming the classical $\text{OMP}_C$ algorithm lies in its method of detecting the support $\textmd{I}_{\vs^\star}$. Another contributing factor is related to the numerical properties of the system matrix, including the concept of spark and RIP (see equations~\eqref{cs-2} $\&$~\eqref{RIP-1}), and mutual coherence.  In fact, let $\vx = \mQ\vs^\star$ with $\|\vs^\star\|_0\leq \kappa$. It was shown in \cite{donoho2003optimally} 
and \cite{tropp2004greed}  that both the classical $\text{BP}_C$ and $\text{OMP}_C$ algorithms can yield the true solution $\vs^\star$ from the system $\vx =\mQ \vs$ with $\mQ$ (column) $l_2$-normalized, if the following condition holds:
\e \kappa < \frac{1}{2}(1 + \frac{1}{\mu[\mQ]}),\label{condi-4-BP-OMP}\ee
where $\mu[\mQ]$ is the
mutual coherence of  $\mQ \in \mathbb R^{N\times L}$.  As shown in \cite{strohmer2003grassmannian}, for a matrix $\mQ \in \mathbb R^{N\times L}$ with $\|\mQ(:,l)\|_2=1,~\forall~l$,
$\mu[\mQ]\leq \sqrt{\frac{L-N}{N(L-1)}}\triangleq \underline{\mu}$ and the lower bound is achieved if and only if the Gram matrix $\mG_{\mQ}\triangleq \mQ^\top\mQ$ satisfies $\mG_{\mQ}(i,j)=\underline{\mu}, \forall~i\neq j$. Such a $\mQ$ is called an equiangular tight-frame (ETF). 

The condition (\ref{condi-4-BP-OMP}) is just a sufficient condition to ensure a perfect recovery of $\vs^\star$. As shown in \cite{elad2007optimized} and \cite{li2013projection}, the sparse recovery accuracy of $\text{OMP}_C$ is more closely related to averaged mutual coherence of $\mQ$, denoted as $\mu_{av}[\mQ]$, which is defined as the mean of all the squared off-diagonal entries of $\mG_{\mQ}$ rather than $\mu[\mQ]$. This implies that it is desired for the system matrix $\mQ$ to have its  Gram matrix close to the identity matrix $\mI$. Based on this argument, it was shown in \cite{li2013projection} and \cite{chen2013projection} that an optimal $\mQ$ is a tight-frame (TF), whose  Gram matrix $\mG_{\mQ}$ is of form
$\mG_{\mQ}=\mV\MAT{cc}\alpha^2 \mI_N & \mzero\\ \mzero&\mzero\mat \mV^\top,$
where $\mV$ is orthonormal. As ETFs exist only for some pairs $(N,L)$, tight-frames provide an alternative for designing sparse recovery systems~\cite{tropp2005designing,eldar2002optimal,tsiligianni2014construction}.

Our proposed $\text{Alg}_{GL2}$ algorithm is based on the system $\vs_0(\vx)=\bar \mQ \vs$. As shown in 
(\ref{HD-system-0}), $\bar \mQ = \mV_1\mT_0 \mQ$. We note that 
\begin{itemize}
\item It is difficult to find an explicit relationship between $\mu[\bar \mQ]$ and $\mu[\mQ]$. However, $\bar \mQ= \mV_1\mV^\top_1$ is actually a 1-tight frame and extensive examples have shown that $\mu_{av}[\bar \mQ]< \mu_{av}[\mQ]$;

\item Let $\mQ$ be a matrix with the $(\kappa, \delta)$-RIP property. It follows from $\mQ=\mU_1\mSigma\mV^\top_1$ that
$$\sigma^{-1}_1\|\mQ\vs\|_2\leq \|\mV^\top_1\vs\|_2\leq |\sigma^{-1}_{\tilde N}\|\mQ\vs\|_2, $$ where $\sigma_1$ and $\sigma_{\tilde N}$ are the smallest and largest (nonzero) singular values of $\mQ$, respectively. It can be shown that with some manipulations that $\bar \mQ=\mV_1\mV^\top_1$ satisfies the $(\kappa, \bar \delta)$-RIP property with
$\bar \delta < \delta+ \frac{(1-\delta)(\sigma_1/\sigma_{\tilde N}-1)}{\sigma_1/\sigma_{\tilde N}+(1-\delta)/(1+\delta)}.$
If $\mQ$ is well-conditioned,\footnote{This is usually a necessary condition for the sparsifying systems to function properly.} i.e., the condition number $\sigma_1/\sigma_{\tilde N}$ is not far away from one, so is $\bar \mQ$. Thus, most of the theoretical results for the original system $\vx = \mQ\vs$ hold for the 
transformed system
$\vs_0(\vx) =\bar \mQ\vs$.
\end{itemize}

The classical $\text{OMP}_C$ algorithm intends to find $(\tilde \vbeta, \mathcal I_k)$ such that the residual  $\vr_k=\vx - \mQ_k\vbeta\triangleq \vx -\mQ \tilde \vs_k$ goes to nil as much as possible. Clearly, $\vr_k = \mQ(\vs^\star - \tilde \vs_k)\triangleq \mQ \Delta\vs$. Assume that $\Delta\vs$ is uncorrelated with a zero-mean. Thus, $\mR \triangleq E[\Delta\vs(\Delta\vs)^\top]$ is diagonal with $\mR(l,l)=\sigma^2_l,~\forall~l$, where $E[.]$ denotes the statistical average operation. Then, $\eta_{av}\triangleq  E[\|\vr_k\|^2_2]$ is given by
$\eta_{av}=\text{Tr}[\mQ^\top\mQ \mR]=\sum^L_{l=1}\sigma^2_l$,
as $\mQ$ is (column) $l_2$-normalized.

Applying the above to the higher-dimensional system $\vs_0=\bar \mQ \vs$, we obtain
$ \bar \eta_{av}=\text{Tr}[\bar\mQ\mR]=\sum^L_{l=1}\|\mV_1(l,:)\|^2_2\sigma^2_l.$   
    Since $\|\mV_1(i_k,:)\|_2\leq 1$, we can see that $\bar \eta_{av} \leq \eta_{av}$. 
    \SL{This establishes the statistical improvement in convergence behavior over $\text{OMP}_C$. Then, we have the following result.}
\begin{proposition}
 \SL{Consider the residual $\vr_k = \mQ \Delta\vs$ with $E[\Delta \vs \Delta (\vs)^\top] = \sigma^2 I$ and
column-normalized $\mQ$, the average residual energies of $\text{Alg}_{GL2}$ and $\text{OMP}_C$ satisfy
\[
\bar \eta_{av} \le \eta_{av}.
\] 
Thus, $\text{Alg}_{GL2}$ achieves a smaller expected residual contraction factor compared with $\text{OMP}_C$. }
\end{proposition}

Now, let us consider the implementation complexity of  $\text{Alg}_{GL2}$. Define
$\mQ_c\triangleq \mV^\top_1.$
It follows from (\ref{characterization}) and $\bar \mQ=\mV_1\mV^\top_1$ that  the 1st equation of (\ref{Greedy-Approach-OMP-HD}) can be rewritten as
$$\|\vs_0 - \bar \mQ(:,\mathcal I_k)\vbeta\|^2_2 = \|\vx_c - \mQ_c(:,\mathcal I_k)\vbeta\|^2_2.$$
Comparing to its counterpart in $\text{OMP}_C$: $\|\vx - \mQ(:,\mathcal I_k)\vbeta\|^2_2$, we can see that at the stage of obtaining the optimal coefficients,  our proposed  $\text{Alg}_{GL2}$ has exactly the same computation complexity if $\vx_c$ is available. We note that the extra computational burden, denoted as $\mathcal C_b$, for computing $\vx_c=\mT_0\vx$  involves $N^2$ multiplications and $N(N-1)$ additions.

Regarding the index selection stage, i.e., the 2nd equation of (\ref{Greedy-Approach-OMP-HD}), our proposed $\text{Alg}_{GL2}$ algorithm needs to compute the residual as follows
$$\hat \vs^\star_k = \vs_0 - \bar \mQ(:,\mathcal I_k)\vbeta = \mV_1(\vx_c-\mQ_c(:, \mathcal I_k)\vbeta),$$
while $\text{OMP}_C$ requires computing $\vxi$ with
$$\vxi = \mQ^\top\vr_k = \mQ^\top(\vx - \mQ(:,\mathcal I_k)\vbeta).$$
This implies that both algorithms have the same computational complexity at this stage. 

Therefore, we claim that our proposed $\text{Alg}_{GL2}$ algorithm significantly outperforms  $\text{OMP}_C$, albeit at the price of an extra computational burden $\mathcal C_b$.\footnote{Note that converting the original system $\vx = \mQ\vs$ into the higher dimensional system $\vs_0(\vx)=\bar \mQ\vs$ requires the SVD of $\mQ$. Since this regards designing the system/algorithm and is done once only, the computational complexity involved at this stage is not considered part of the algorithm's overall complexity.}

\subsection{Comparison between $\text{BP}_C/\text{BP}_{alt}$ and $\text{Alg}_{GL1}$} \label{sec:comp_BP}
Recall that $\text{BP}_C$ refers to the BP algorithm based on the classical formulation (\ref{BP-1x}), while $\text{BP}_{alt}$ denotes the BP algorithm based on the reformulation (\ref{alternative-2x}) of (\ref{BP-1x}). They are mathematically equivalent.  

Compared to $\text{OMP}_C$, the two BP-based algorithms adopt a global optimization approach to detect $\textmd{I}_{\vs^\star}$. However, the interactions among the entries in $\vs$, imposed by the constraint $\vx =\mQ\vs$, make it challenging to identify the sparsest solution $\vs^\star$. The proposed $\text{Alg}_{GL1}$ algorithm is 
specifically designed to mitigate the influence of the magnitudes of detected atoms on the identification of subsequent indices in a greedy manner. It leverages the strengths of both the classical $\text{BP}_{alt}$ and $\text{OMP}_C$ algorithms, effectively combining their advantages. Consequently, the substantial performance improvement over   $\text{BP}_C/\text{BP}_{alt}$, as previously discussed, primarily stems from the method used to detect the support $\textmd{I}_{\vs^\star}$, which is performed in a higher-dimensional space.

The proposed $\text{Alg}_{GL1}$ algorithm exhibits excellent performance in terms of sparse recovery accuracy. Regarding implementation complexity, however, it requires solving $\kappa$ $l_1$-minimization problems of the form $\min_{\vz}\|\vs_0 +\bar \mW \vz\|_1$, i.e., \eqref{alternative-2x}, using $\text{BP}_{alt}$. 
\SL{
Therefore, the computational cost of $\text{Alg}_{GL1}$ is generally higher than that of a single run of $\text{BP}_{C}$/$\text{BP}_{alt}$, as $\text{Alg}_{GL1}$ applies $\text{BP}_{alt}$ iteratively to select one index at a time. This reflects the expected tradeoff between improved recovery accuracy and increased computation, as 
demonstrated in Tables~\ref{tab:recovery_rates_horizontal}-\ref{table-SZ-1y} and Fig.~\ref{fig_kappa_varying_N64}. }

More efficient algorithms for solving (\ref{alternative-2x}) are needed to enable the use of $\text{Alg}_{GL1}$ in large-scale sparse recovery systems. 
Indeed, a subgradient-based algorithm was proposed in~\cite{li5069675revisiting} 
for solving \eqref{alternative-2x}, which significantly outperforms $\text{BP}_{alt}$ in terms of speed and has substantially lower implementation complexity. In the next section, we introduce an IRLS-based algorithm.
By replacing $\text{BP}_{alt}$ with this IRLS-based approach in $\text{Alg}_{GL1}$, the resulting algorithm not only achieves significantly faster convergence but also delivers superior sparse recovery performance.

\section{Improved Algorithms}\label{sec-4}
The proposed $\text{Alg}_{GL1}$ algorithm is computationally slow due to two main factors: i) it relies on linear programming-based BP; and ii) similar to the classical OMP, it selects atoms one-by-one. Additionally, the proposed algorithm $\text{Alg}_{GL1}$ is
designed to solve the minimization problems of the form \eqref{Greedy-Approach} with $\chi$ defined as the $l_1$-norm. Higher performance algorithms can be developed by employing a more effective norm to approximate the sparsity norm $\|.\|_0$.

In this section, we derive a class of efficient algorithms that adopt an atom update scheme similar to that of CoSaMP and explore strategies for handling cases where the measurements are contaminated by noise.

\subsection{Iteratively re-weighted approach to \eqref{Greedy-Approach}}
Note that the cost function  $F(\vz,\vbeta)$  of \eqref{Greedy-Approach} is of form
\eqn f(\bar \vz)&\triangleq &\|\vs_0 + \mOmega\bar \vz\|_{\chi}~~~~~~~~\text{s.t.}\quad \mOmega =\vs_0 + \MAT{cc}\mW & -\bar \mQ_k\mat, \bar \vz = \MAT{c}\vz\\ \vbeta\mat, \nonumber 
\eeqn
where $\chi$ represents an approximating function $\phi(t)$ of the sign function $\text{sgn}[|t|]$. 

Thus, the key to \eqref{Greedy-Approach} is to solve
\e \min_{\bar \vz} f(\bar \vz)=\min_{\bar \vz}\sum_l 
\phi(\vs_0(l)+\mOmega(l,:)\bar \vz).\label{General-cost-F}\ee
Here, we consider two approximating functions defined before:
$\phi_1(t)=|t|, ~~\phi_3(t)=(|t|^2+\epsilon)^{q/2},$
where $\epsilon>0$ and $0<q<1$. When $\chi$ represents $\phi_1(t)=|t|$, (\ref{General-cost-F}) is convex and can be solved using $\text{BP}_{alt}$, resulting in the $\text{Alg}_{GL1}$ algorithm.

%
%
%

When $\chi$ represents the $l_q$-norm $\phi_3(t)$, $f(\bar \vz)$ is highly non-convex and such a problem can be addressed using the IRLS method, which generates a sequence $\{\bar \vz^{(k)}\}$ for a fixed $\epsilon$ in the way:
\eqn \bar \vz^{(k+1)} \triangleq&\arg\min_{\bar \vz}\|\bar \vs_0 + \bar \mOmega \bar \vz\|^2_2, \quad
\vs^{(k+1)} = \vs_0+\mOmega \bar \vz^{(k+1)}, \label{Alt-rw_LS-1x}\eeqn
where $\bar \vs_0\triangleq \mD^{(k)}_2\vs_0$ and $\bar \mOmega \triangleq \mD^{(k)}_2\mOmega$, with the diagonal $\mD^{(k)}_2$ given by
\e  \mD^{(k)}_2(l,l) = \sqrt{[|\vs^{(k)}(l)|^2+\epsilon]^{q/2-1}},~\forall~l.\label{weightings}\ee
Note that the solution to the 1st term of \eqref{Alt-rw_LS-1x} is given by
\e \bar \vz^{(k+1)} =-(\bar \mOmega^\top\bar\mOmega)^{-1}\bar \mOmega^\top\bar \vs_0.\label{Alt-rw_LS-2x}\ee
We use the same $\epsilon$-regularization scheme to set $\{\epsilon_j\}$ and the stopping criterion as in \cite{chartrand2008iteratively}. The corresponding $\epsilon$-regularized $\text{IRLS}$ algorithm, denoted as $\text{IRLS}_{LQ-alt}$ to differentiate the classical $\text{IRLS}_C$, is presented in Algorithm~\ref{alg:IRLS_LQ_alt}. For convenience, we denote this algorithm as
$[\vs, \vz] = \text{IRLS}_{LQ-alt}(\vs_0,\mOmega,q).$ 
As understood, with $\text{BP}_{alt}$ replaced by $\text{IRLS}_{LQ-alt}$ in $\text{Alg}_{GL1}$ a similar algorithm to $\text{Alg}_{GL1}$ can be obtained. Such an algorithm is denoted as $\vs^\ast = \text{Alg}_{GLQ}(\vs_0, \mOmega,q)$.

\begin{algorithm}
\caption{An IRLS-based algorithm for solving \eqref{General-cost-F} ($\text{IRLS}_{LQ-alt}$)}\label{alg:IRLS_LQ_alt}
\begin{algorithmic}
\State\textbf{Inputs:} $(\vs_0, \mOmega, q)$;
\State\textbf{Initials:} 
Set  $\epsilon=10$, $\vz=\vzero$, $\vs = \vs_0$, and $\vs_p=\vzero$;
\For {$\tilde p=1,2, \cdots,9$} 
\State (1) set $\epsilon=\epsilon/10$ and $ct=0$;
\State (2) while $\|\vs-\vs_p\|_2/\|\vs\|_2 > \sqrt{\epsilon}/100$, set $\vs_p=\vs$ and $ct ~\leftarrow~ct+1$. Then, compute the weighting matrix $\mD_2$ with \eqref{weightings};
\State (3) update $\vs$ with
$ ~~\vz = -(\mOmega^\top\mD_2^2\mOmega)^{-1}\mOmega^\top\mD_2^2\vs_0,\quad
\vs = \vs_0 +\mOmega \vz;\nonumber $ 
\State (4) check~~
   ${\bf If} ~ct > 4500, ~{\bf break}.$
\State (5)  Go to step (2);
\EndFor
\State\textbf{Output:} $\vs$.
\end{algorithmic}
\end{algorithm}

\begin{Remark}\label{Remark 4.1}
    It should be noted that the solution to \eqref{Alt-rw_LS-1x} is theoretically obtained through matrix inversion as shown in~\eqref{Alt-rw_LS-2x}.  However, this solution for the unconstrained LS problem can also be computed using more efficient iterative algorithms, such as GP and CGP, which avoid matrix inversion. See \cite{blumensath2008gradient} for details. This provides a significant advantage of our proposed $\text{IRLS}_{LQ-alt}$ over the classical $\text{IRLS}_C$, which requires a large number of matrix inversions (see \eqref{matrix-inv-LS-cla}). These inversions are not only computationally expensive but can also lead to numerical issues when the system matrix $\mQ$ is ill-conditioned. This advantage will be further demonstrated in the simulation section.
\end{Remark}

\subsection{CoSaMP-type algorithms for sparse recovery }
As mentioned earlier, although $\text{Alg}_{GL1}$ achieves excellent sparse recovery performance, it is very slow, partially due to the one-by-one update of atoms. By using the same idea as in CoSaMP, we can update the set of selected atoms with multiple atoms at each iteration.

Let $\text{Alg}_{GLX}$ be an algorithm, say $\text{Alg}_{GL1}$ or $\text{Alg}_{GLQ}$, derived earlier. Its improved version, which uses a CoSaMP-type strategy for updating the set of selected indices, is denoted as $\text{Alg}^F_{GLX}$ and outlined in Algorithm~\ref{alg:FGL1} (for $\text{Alg}^F_{GL1}$). For convenience, we denote this algorithm as
$\vs^\ast = \text{Alg}^F_{GL1}(\vs_0, \mW, \mV_1,\kappa,\kappa_p).$
By replacing $\text{BP}_{alt}$ with $\text{IRLS}_{LQ-alt}$ in $\text{Alg}^F_{GL1}$, we obtain $\text{Alg}^F_{GLQ}$.

\begin{algorithm}
\caption{A CoSaMP-type sparse recovery algorithm associated with $\text{BP}_{alt}$ ($\text{Alg}^F_{GL1}$)}\label{alg:FGL1}
\begin{algorithmic}
\State\textbf{Inputs:} $(\vs_0, \mW, \mV_1)$, $\kappa$, and  $\kappa_p$;
\State\textbf{Initials:} Run
$ [\tilde \vz,\tilde \vs]=\text{BP}_{alt}(\vs_0, \mW),$
Then, assign the residual $\tilde \vs$ to $\vs$, and set $\mathcal I$ as the set of indices for the $\kappa_p$ largest entries of $\vs$ in magnitude;
\For {$k=1:N_{iter}$}~~
\State - Checking whether $\vs$ is the desired solution: Let $|\vs|_{\kappa+1}$ denote the $(\kappa+1)$-th entry of the  re-arranged sequence of $\{|\vs(l)|\}$ sorted in a descending order.
$~~{\bf If} ~~|\vs|_{\kappa+1} \leq 10^{-4}, ~ {\bf break}~~ {\it for-loop}.$
\State - Updating the residual $\tilde \vs$ and $\tilde \vbeta$: Compute $\mOmega = \MAT{cc}\mW& -\mV_1(\mV_1(\mathcal I,:))^\top\mat$, then run  $[\vz,\tilde \vs] = \text{BP}_{alt}(\vs_0,\mOmega)$.
Set $\tilde \vbeta = \vzero$ and $\tilde \vbeta(\mathcal I)$ with the last $\text{length}(\mathcal I)$ entries of $\vz$;
\State - Updating $\vs$ using  (\ref{Insert-X}) with the pair $(\tilde \vs, \tilde \vbeta)$:~~
$\vs = \tilde \vs +  \tilde \vbeta.$
\State - Updating the index set $\mathcal I$ with the indices of the $\kappa$ largest entries of $\vs$ in magnitude.
\EndFor
\State\textbf{Output:} $\vs^\ast=\vs$.
\end{algorithmic}
\end{algorithm}

\begin{Remark} \label{Remark 4.2}
We note that (i)
the parameter $\kappa_p$ is determined by the sparse recovery performance of the associated algorithm, like $\text{BP}_{alt}$ and $\text{IRLS}_{LQ-alt}$. In general, $\kappa_p$ is chosen such that  $0.35\leq \kappa_p/\kappa \leq 0.75$;
(ii) as observed, the number of indices selected in $\mathcal I$ is equal to $\kappa$, except during the first iteration, where $\kappa_p$ indices are included. This number can be slightly greater than $\kappa$ but is definitely smaller than $2\kappa$. 
See Theorem~\ref{theorem-Greedy-Alt}.
\end{Remark}

\subsection{Robust sparse recovery algorithms against low-rank disturbance}
\label{sec:robust}
So far, our discussion has been limited to the case where the measurement signal $\vx$ is strictly sparse. Now, we consider a more practical scenario where the signal $\vx$ is of the following form
\e \vx = \mQ\vs + \ve\triangleq \vx_0 +\ve,\label{restart-2}\ee
where $\ve$ represents a disturbance that affects the recovery of the true sparse vector $\vx_0$ from $\vx$. 

The objective here is to separate $\vx_0$ and $\ve$ from a $\vx$ given. A traditional approach to address this problem is to first obtain a primary estimate $\vs^\ast$ of the true $\vs^\star$ underlying $\vx_0$ by applying a sparse recovery algorithm to $\vx$. Then, defining $\mathcal I$ as the set of indices corresponding to the $\kappa$ largest entries of $\vs^\ast$ in magnitude, an estimate $\hat \vx_0$ of the clean  $\vx_0$ is given by  
\e \hat \vx_0 = \mQ \hat\vs^\star,\label{classical-Oracle}\ee
where $\hat\vs^\star$ is  $\kappa$-sparse and satisfies $\hat\vs^\star(\mathcal I)=\vbeta^\ast$, with
$\vbeta^\ast\triangleq \arg\min_{\vbeta}||\vx-\mQ(:,\mathcal I)\vbeta||^2_2.$
This estimation procedure corresponds to the classical Oracle estimator \cite{candes2005decoding} when $\mathcal I = \textmd{I}_{\vs^\star}$.  In general, the classical sparse recovery algorithms are highly sensitive to noise $\ve$. Even when $\mathcal I = \textmd{I}_{\vs^\star}$, the obtained estimate $\hat \vx_0$ does not necessarily equal $\vx_0$ unless $\ve = \vzero$.

There are scenarios, in which $\ve$ is often modeled as a low-rank signal \cite{dao2014structured,tivive2018gpr}:
\e \ve = \mGamma \valpha^\star,\label{restart-3}\ee 
where $\mGamma \in \mathbb R^{N\times N_e}$ with  $N_e<< N$ is a matrix satisfying $\mGamma^\top\mGamma=\mI_{N_e}$, and $\valpha^\star$ is an unknown vector. In multiple measurement vector models, \eqref{restart-2} transforms into a low-rank and sparse decomposition problem, a framework that has been extensively studied in the literature~\cite{candes2011robust,zhou2011godec,otazo2015low,bahmani2016near,wright2013compressive}. Similar to the Robust PCA (RPCA)-based approach introduced in~\cite{candes2011robust}, most of these works focus on the special case where $\mQ = \mI_N$, investigating various methods for separating the low-rank and sparse components. In this paper, we address the problem of recovering the clean signal $\vx_0$ from  $\vx$ as given in \eqref{restart-2}, with $\ve$ modeled by \eqref{restart-3}. We consider the scenario where both $\mQ$ and $\mGamma$ are known, with $\mQ$ having dimensions $N\times L$ and assuming $N_e<< N\leq L$.



Note $\vx-\ve = \mQ\vs$. Under our proposed characterization, we have $\vs = \mT_0(\vx-\ve)+\mW\vz$, and hence
\e\vs =\mV_1\mT_0\vx +\MAT{cc}\mW&-\mV_1\mT_0\mGamma\mat \MAT{c}\vz\\ \valpha\mat\triangleq \vs_0 + \mW_c \vz_c.\nonumber\ee  
Based on this formulation, we propose the following robust approach to sparse recovery 
\eqn \vz^\star_c &\triangleq& \arg\min_{\vz_c}\|\vs_0+\mW_c\vz_c\|_0\quad\mapsto ~~~~\vs^\star = \vs_0+\mW_c\vz^\star_c.\label{restart-4} \eeqn
Let $\mW_c \in \mathbb R^{L\times L_c}$. It was shown in~\cite{li5069675revisiting} that \eqref{restart-4} has a unique $\kappa$-solution $\vs^\star$ if and only if all the $(L-2\kappa)\times L_c$ submatrices of $\mW_c$ are of rank $L_c$, which holds when $\mW_c=\mW$. Since, in general, $L_c=L-N+N_e$, ensuring this condition requires that 
$L-2\kappa \geq L_c$, which simplifies to 
$2\kappa + N_e \leq N.$

As observed, (\ref{restart-4}) has the same form as (\ref{restart-1}). Consequently, it can be addressed using classical methods such as $\text{BP}_{alt}$ and the just derived $\text{IRLS}_{LQ-alt}$. More interestingly, with $(\mW, \vz)$ replaced by $(\mW_c, \vz_c)$, our proposed greedy algorithms such as $\text{Alg}_{GL1}$, $\text{Alg}_{GLQ}$, $\text{Alg}^F_{GL1}$ and $\text{Alg}^F_{GLQ}$, can also be applied to solve \eqref{restart-4}.

\begin{Remark} \label{Remark 4.3}
It is important to highlight the following key points: 
\begin{itemize}\item The results obtained above indicate that the true support $\textmd{I}_{\vs^\star}$ can be recovered with high probability using our proposed algorithms, even in the presence of disturbance $\ve$. This demonstrates a significant advantage of our proposed sparse model $\vs =\vs_0(\vx) + \mW\vz$ over the classical formulation $\vx = \mQ\vs$;
\item Let $\vs^\ast$ be the primary estimate of $\vs^\star$ obtained using one of our proposed algorithms. Suppose that $\mathcal I$ - the set of indices corresponding to the $\kappa$ largest entries of $\vs^\ast$, is equal to $\textmd{I}_{\vs^\star}$. In this case, the estimate $\hat\vx_0$ can be computed using the classical Oracle estimator \eqref{classical-Oracle}. Alternatively, it follows from $\vs^\star = \vs_0+\mW_c\vz^\star_c$ and $\vs^\star(\mathcal I)=\vzero$ that 
$\vzero = \vs_0(\mathcal I) +\mW_c(\mathcal I,:)\vz^\star_c.$
Thus,  we have 
\e \vz^\star_c = -((\mW^\star_c)^\top\mW^\star_c)^{-1}(\mW^\star_c)^\top\vs_0(\mathcal I)=\MAT{c}\vz^\star\\ \alpha^\star\mat, \nonumber
\ee
where  $\mW^\star_c\triangleq \mW_c(\mathcal I,:)$. The corresponding estimate of $\vx_0$ is given by 
$\hat \vx_0 = \vx - \mGamma \valpha^\star = \vx_0.$
This result implies that the clean signal $\vx_0$ can be exactly recovered using the framework based on our proposed model or characterization, provided that $\mathcal I=\textmd{I}_{\vs^\star}$;
\item Note $\vx = \mQ\vs + \mGamma \valpha =\MAT{cc}\mQ&\mGamma\mat\MAT{c}\vs\\ \valpha\mat\triangleq \tilde \mQ \tilde \vs$. The sparsity-based signal separation problem can then be formulated as the classical form:
$$\tilde\vs^\star\triangleq \arg\min_{\tilde \vs}||\mD_0\tilde \vs||_0~~~\text{s.t.}~~\vx=\tilde \mQ \tilde \vs,$$
where $\mD_0\triangleq \MAT{cc}\mI_L&\mzero\\ \mzero&\mzero\mat$. However, solving this problem using classical algorithms such as $\text{OMP}_C$ and $\text{IRLS}_C$ is challenging due to the potential numerical ill-conditioning of $\tilde \mQ$. In particular, $\text{IRLS}_C$ can not be directly applied to the problem due to the singularity of $\mD_0$. Furthermore, even if the true support $\textmd{I}_{\vs^\star}$ is correctly identified using a classical algorithm, the clean signal $\vx_0$ cannot be recovered using the classical Oracle estimator, as previously discussed. 
\end{itemize}
\end{Remark}

\section{Numerical examples and simulations}\label{sec-5}
In this section, we present a number of numerical examples and simulations in MATLAB to demonstrate the performance of five classical algorithms and five proposed algorithms:
\begin{itemize}
\item $\text{OMP}_C$ -the classical orthogonal matching pursuit (OMP) algorithm~\cite{mallat1993matching,tropp2004greed};
\item $\text{CoSaMP}_C$ - the classical compressive sampling match
pursuit (CoSaMP) algorithm~\cite{needell2009cosamp};
\item $\text{BP}_{C}$  - the classical basis pursuit (BP) algorithm~\cite{chen2001atomic,candes2005decoding};
\item $\text{IRLS}_C$ - the iteratively re-weighted least squares (IRLS) algorithm proposed in  \cite{chartrand2008iteratively} with $q=1/2$;
\item $\text{FISTA}$  - the fast iterative shrinkage-thresholding (FISTA) algorithm proposed in~\cite{beck2009fast};
\item $\text{Alg}_{GL2}$ - the proposed greedy algorithm using $l_2$;
\item $\text{Alg}_{GL1}$  - the proposed greedy algorithm with atoms selected one-by-one using $\text{BP}_{alt}$;
\item $\text{Alg}^F_{GL1}$ - the proposed fast greedy algorithm with $\kappa$ atoms selected at each iteration using $\text{BP}_{alt}$;
\item $\text{Alg}_{GLQ}$ - the proposed greedy algorithm with atoms selected one-by-one using $\text{IRLS}_{LQ-alt}$;
\item $\text{Alg}^F_{GLQ}$ - the proposed fast greedy algorithm with $\kappa$ atoms selected at each iteration using $\text{IRLS}_{LQ-alt}$.
\end{itemize}

\subsection{Data setting and measures}
The synthetic data used in this section is generated as follows: For a given {\it setting} $(\kappa, N, L)$, we create a matrix of dimension $N\times L$ using {\it randn}, and the system matrix $\mQ$ is obtained by normalizing the columns of this matrix. A sequence of $J$ $\kappa$-sparse samples, $\{\vs_j\}$, is generated with the positions of the non-zero entries randomly selected using {\it randperm}, while these entries are produced using {\it randn} such that $||\mQ\vs_j||_2=1,~\forall ~j$. 

To evaluate the performance of a sparse recovery algorithm, we adopt the rate of successful sparse recovery $ \varrho_{ok}$ and the signal-to-noise ratio (SNR), both defined at the end of Section~\ref{sec-2}. Furthermore, the {\it wall-clock} time in seconds, denoted as $T_c$, represents the total time taken by an algorithm to process the $J$ samples.

\subsection{Sparse recovery accuracy and computational efficiency}

\begin{table}[thb!]\caption{Statistics of the rate $\varrho_{ok}$ and $T_c$ of the ten algorithms for different settings $(\kappa, N, L)$ with $J=100$.}\label{table-SZ-1y}
\centering
\smaller
\setlength{\tabcolsep}{3pt}
\renewcommand{\arraystretch}{1.02}
\resizebox{\linewidth}{!}{%
\begin{tabular}{c|c||c|c|c|c|c|c|c|c|c|c}
\hline \cline{1-12}  \hline \cline{1-12}

{\tiny$ \MAT{c}\kappa\\N\\ L\mat$} & &$\text{OMP}_C$& $\text{CoSaMP}_C$& $\text{BP}_C$& $\text{IRLS}_C$&$\text{FISTA}$&$\text{Alg}_{GL2}$&$\text{Alg}_{GL1}$&$\text{Alg}^F_{GL1}$&$\text{Alg}_{GLQ}$&$\text{Alg}^F_{GLQ}$\\ \cline{1-6}  \hline \cline{1-12}
\renewcommand{\arraystretch}{0.99}
\multirow{2}{*}{{\tiny$ \MAT{c}10\\64\\ 128\mat$} } 
& $\varrho_{ok}$  & 99\% & 100\%&100\% & 100\%    &99\%    &100\%    &100\%    &100\% &    100\%   &   100\%   \\ \cline{2-12}
& $T_c$&3.05$\text{e}$-2 &2.09$\text{e}$-2  &8.43$\text{e}$-1& 3.20$\text{e}$-1  &   4.04$\text{e}$-1    &2.22$\text{e}$-2    &2.00$\text{e}$+0  &2.00$\text{e}$+0    &2.26$\text{e}$-1   & 2.06$\text{e}$-1\\ \cline{1-12}

\renewcommand{\arraystretch}{0.99}
\multirow{2}{*}{ {\tiny $ \MAT{c}30\\64\\ 128\mat$} }
& $\varrho_{ok}$  & 6\% &0\%&11\% & 92\%    &3\%    & 53\%    &  94\%    &78\% &    99\%   &94\%   \\ \cline{2-12}
& $T_c$   &2.12$\text{e}$-1 &  9.87$\text{e}$+0& 1.12$\text{e}$+0&1.05$\text{e}$+0  &2.75$\text{e}$+0& 2.27$\text{e}$-1&3.02$\text{e}$+1  &7.34$\text{e}$+0&6.31$\text{e}$+0& 1.87$\text{e}$+0\\ \cline{1-12}
\renewcommand{\arraystretch}{0.99}
\multirow{2}{*}{{\tiny$ \MAT{c}35\\128\\256\mat$}}
& $\varrho_{ok}$  &85\%&100\% &100\% &100\%  &62\%    &100\%    & 100\%    &100\% &100\%&100\%   \\ \cline{2-12}
& $T_c$   & 3.88$\text{e}$-1 &1.76$\text{e}$-1& 4.71$\text{e}$+0& 2.84$\text{e}$+0  &1.91$\text{e}$+1&5.40$\text{e}$-1& 1.47$\text{e}$+1  &1.47$\text{e}$+1&2.32$\text{e}$+0 & 2.29$\text{e}$+0\\ \cline{1-12}
\renewcommand{\arraystretch}{0.99}
\multirow{2}{*}{{\tiny$ \MAT{c}58\\128\\256\mat$}}
& $\varrho_{ok}$  & 8\% &0\%& 4\% & 100\%    &0\%    &56\%    &  100\%    &99\%&  100\%   &100\%   \\ \cline{2-12}
& $T_c$  & 1.62$\text{e}$+0 & 8.14$\text{e}$+1&5.99$\text{e}$+0&5.24$\text{e}$+0 & 2.07$\text{e}$+1&2.38$\text{e}$+0&3.02$\text{e}$+2 &3.58$\text{e}$+1&3.92$\text{e}$+0 &3.81$\text{e}$+0\\ \cline{1-12}

\renewcommand{\arraystretch}{0.99}
\multirow{2}{*}{{\tiny$ \MAT{c}60\\256\\512\mat$}}
& $\varrho_{ok}$  & 90\% &100\%& 100\% & 100\%    &35\%    & 100\%    &  100\%    &100\%&  100\%   &100\%   \\ \cline{2-12}
& $T_c$  &2.68$\text{e}$+0 &  7.78$\text{e}$-1&3.13$\text{e}$+1&9.07$\text{e}$+0 &5.55 $\text{e}$+1&3.69$\text{e}$+0&1.18$\text{e}$+2 &1.18$\text{e}$+2&6.93$\text{e}$+0 &6.77$\text{e}$+0\\ \cline{1-12}

\renewcommand{\arraystretch}{0.99}
\multirow{2}{*}{{\tiny$ \MAT{c}100\\256\\512\mat$}}
& $\varrho_{ok}$  & 16\% &0\%& 53\% & 100\%    &0\%    & 90\%        & 100\%    &100\%&  100\%   &100\%   \\ \cline{2-12}
& $T_c$  & 9.05$\text{e}$+0 &2.82$\text{e}$+2& 4.63$\text{e}$+1&1.44$\text{e}$+1
  &5.90$\text{e}$+1& 1.09$\text{e}$+1&1.03$\text{e}$+3 & 2.05$\text{e}$+2&1.04$\text{e}$+1 &1.04$\text{e}$+1\\ \cline{1-12}
\end{tabular}
}
\end{table}

\begin{figure}[htb!]
\begin{minipage}{0.49\linewidth}
\centering
\includegraphics[width=2.3in]{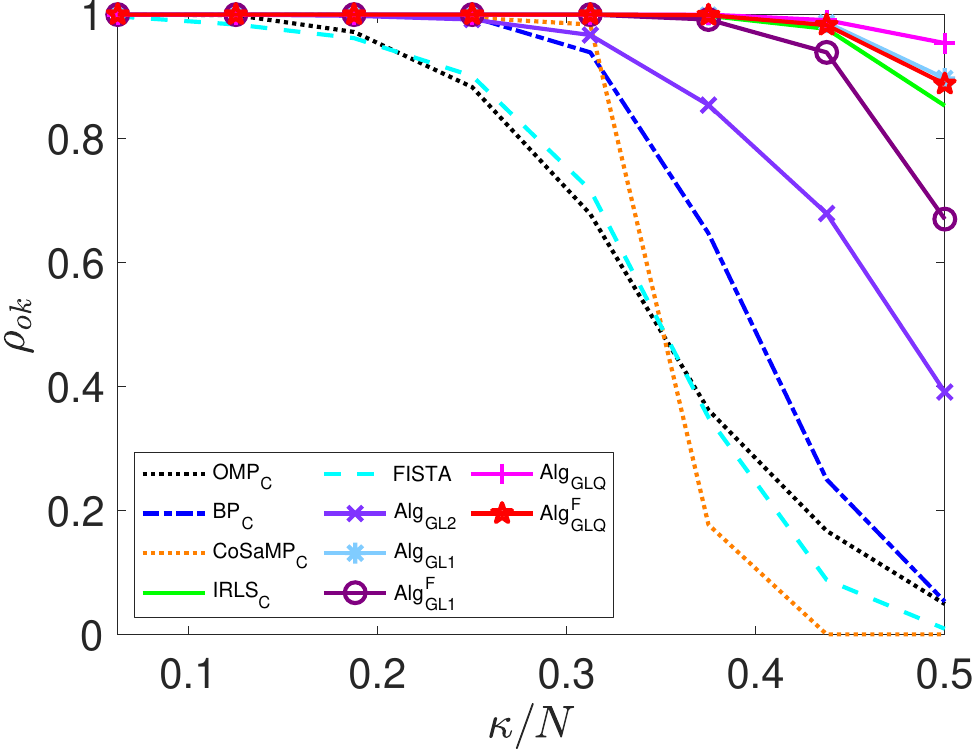}
\centerline{\footnotesize{(a)}}
\end{minipage}
\hfill
\begin{minipage}{0.49\linewidth}
\centering
\includegraphics[width=2.3in]{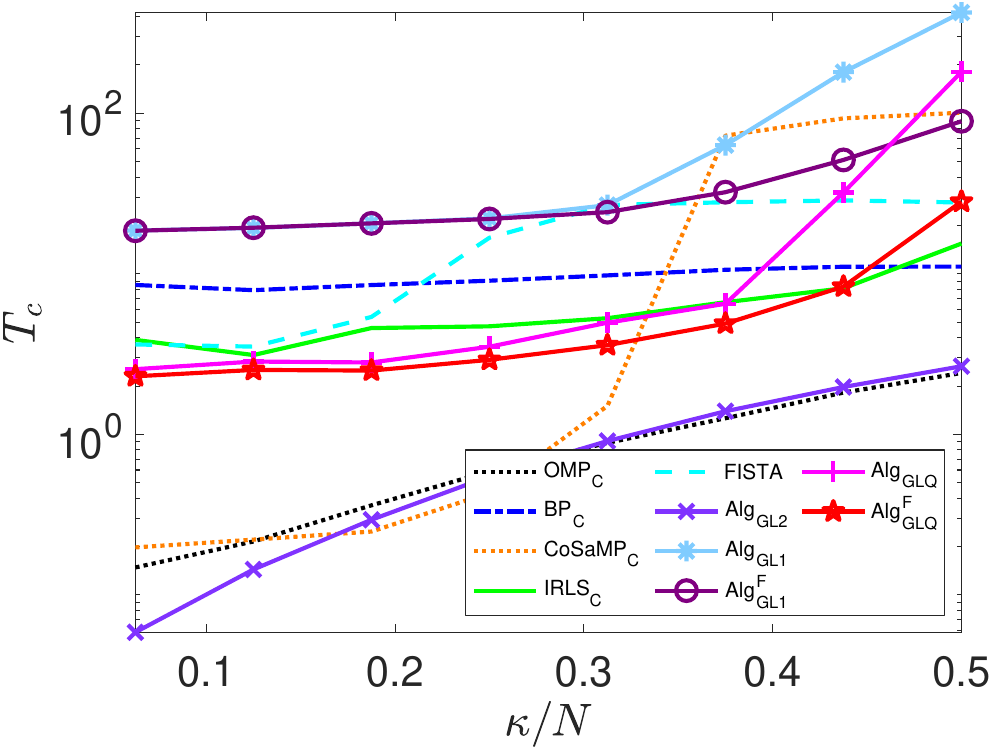}
\centerline{\footnotesize{(b)}}
\end{minipage}
\caption{The rate $\varrho_{ok}$ of successful recovery and the wall-clock time $T_c$  with respect to different $\kappa$ for each of the ten algorithms for $N=64, L=128, J=1000$.}
\label{fig_kappa_varying_N64}
\end{figure}

Table~\ref{table-SZ-1y} shows the performance of the ten algorithms for different settings $(\kappa, N, L)$ with $J=100$ samples.
The ten algorithms are further evaluated under the setting $N=64, L=128$ and $ J=1000$, with $\kappa$ varying from 4 to $N/2=32$ in steps of 4. Fig.~\ref{fig_kappa_varying_N64} shows the corresponding statistics of $\varrho_{ok}$ and $T_c$. 
We note from the simulations above that
1) among the five classical algorithms, $\text{IRLS}_C$ achieves the best sparse recovery performance and is significantly faster than $\text{BP}_C$;
2) the proposed $\text{Alg}_{GL2}$ achieves the same sparse recovery accuracy as $\text{IRLS}_C$ when $\kappa/N$ is small, but with much higher
computational efficiency;
3) the other four proposed algorithms, i.e.,  $\text{Alg}_{GL1}$, $\text{Alg}_{GLQ}$, $\text{Alg}^F_{GL1}$, and $\text{Alg}^F_{GLQ}$, demonstrate a similar performance to $\text{IRLS}_C$. Notably, $\text{Alg}_{GL1}$, $\text{Alg}_{GLQ}$ and $\text{Alg}^F_{GLQ}$ even outperform the state-of-the-art algorithm $\text{IRLS}_C$ in terms of $\varrho_{ok}$.


\begin{figure}[htb!]
\begin{minipage}{0.49\linewidth}
\centering
\includegraphics[width=2.3in]{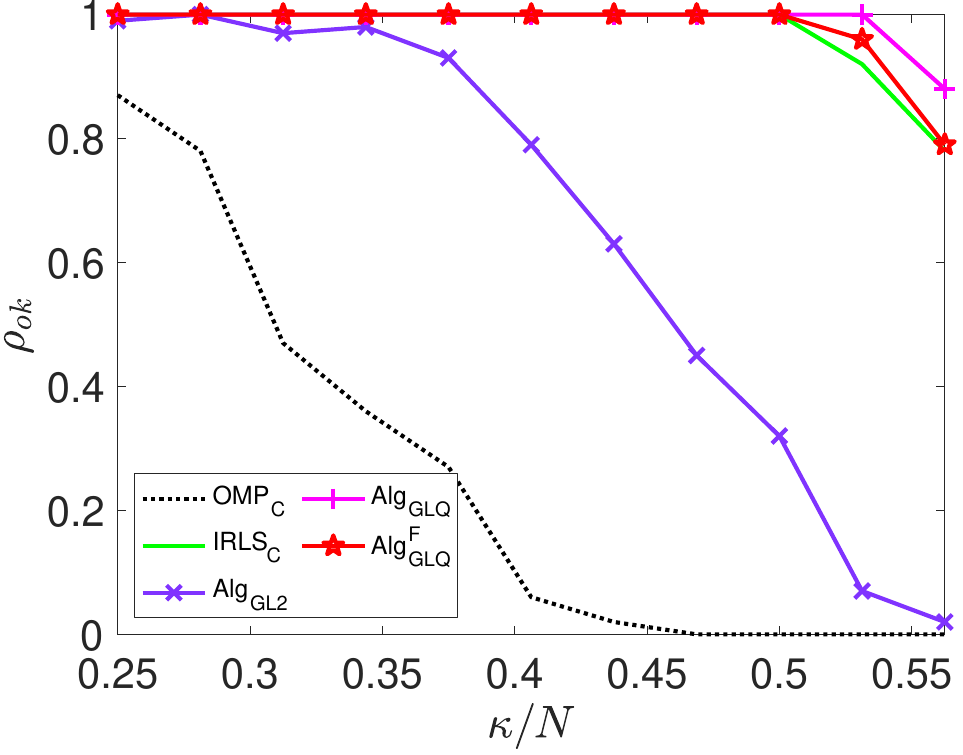}
\centerline{\footnotesize{(a)}}
\end{minipage}
\hfill
\begin{minipage}{0.49\linewidth}
\centering
\includegraphics[width=2.3in]{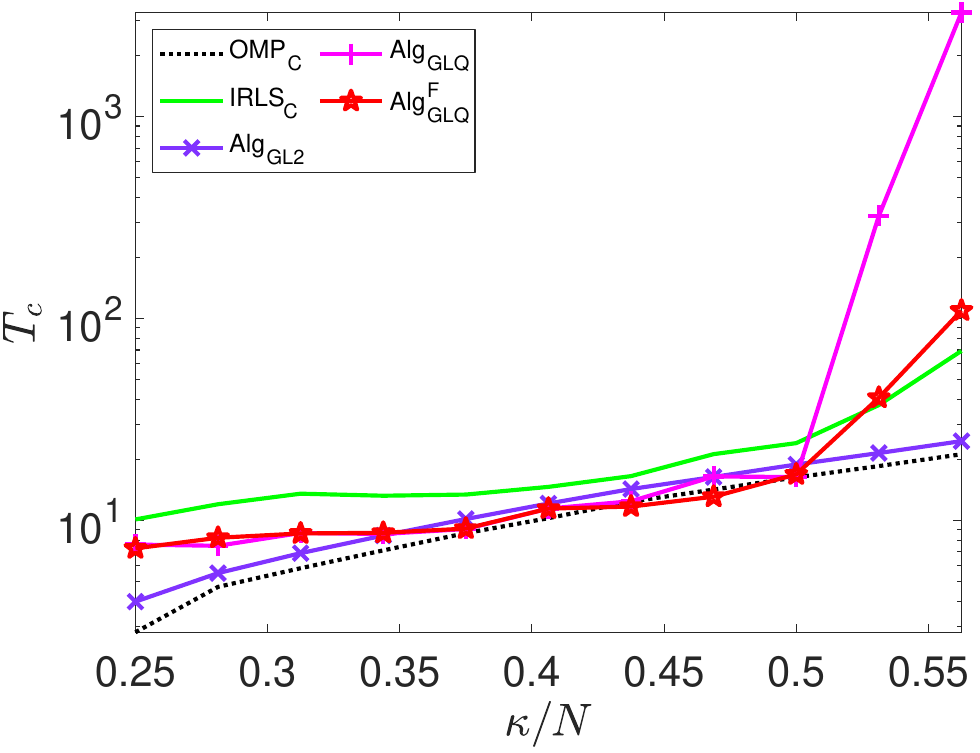}
\centerline{\footnotesize{(b)}}
\end{minipage}
\caption{The rate $\varrho_{ok}$ of successful recovery and the wall-clock time $T_c$  with respect to different $\kappa$ for each of the five algorithms for $N=256, L=512,J=100$. 
} \label{fig_kappa_varying_N256}
\end{figure}

As observed in Table~\ref{table-SZ-1y}, the classical algorithms $\text{BP}_C$,  $\text{CoSaMP}_C$ and $\text{FISTA}$,  along with the proposed $\text{Alg}_{GL1}$ and $\text{Alg}^F_{GL1}$, become very time-consuming as the dimension $(N,L)$ of the system (matrix) $\mQ$ gets large. Therefore, in the sequel, we will focus on the other five algorithms when large-scale systems are concerned.
Fig.~\ref{fig_kappa_varying_N256} presents  simulations similar to those in Fig.~\ref{fig_kappa_varying_N64} for the five algorithms: $\text{OMP}_C$,   $\text{IRLS}_C$, $\text{Alg}_{GL2}$, $\text{Alg}_{GLQ}$ and $\text{Alg}^F_{GLQ}$, under the setting $N=256, L=512$ and $ J=100$, with $\kappa$ ranging from 64 to 144 in steps of 8. As observed, the proposed algorithms $\text{Alg}_{GLQ}$ and $\text{Alg}^F_{GLQ}$ can perfectly reconstruct all $J=100$ samples, similar to the performance of the $\text{IRLS}_C$ algorithm, when $\kappa/N$ is small (e.g., $\kappa/N \leq 0.5$), but with much higher
computational efficiency. However, as $\kappa/N$ increases, the proposed $\text{Alg}_{GLQ}$ and $\text{Alg}^F_{GLQ}$ algorithms demonstrate superior sparse recovery performance, albeit with higher computational time requirements.


\begin{figure*}[htb!]
\begin{minipage}{0.3\linewidth}
\centering
\includegraphics[width=1.6in]{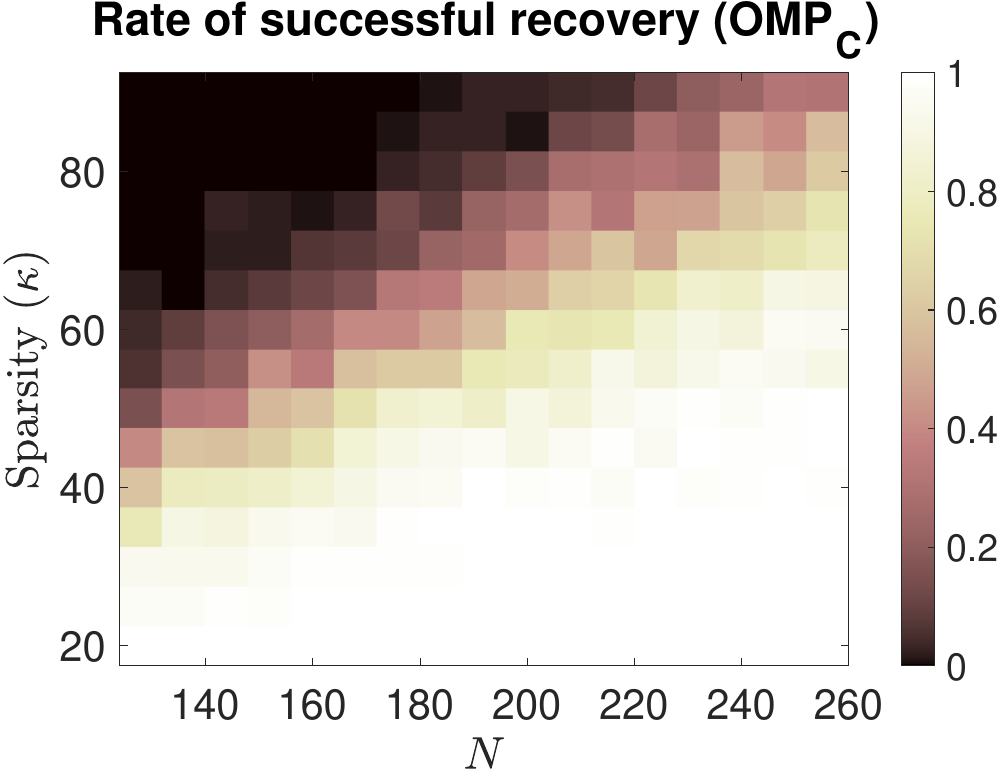}
\centerline{\footnotesize{(a)}}
\end{minipage}
\hfill
\begin{minipage}{0.3\linewidth}
\centering
\includegraphics[width=1.6in]{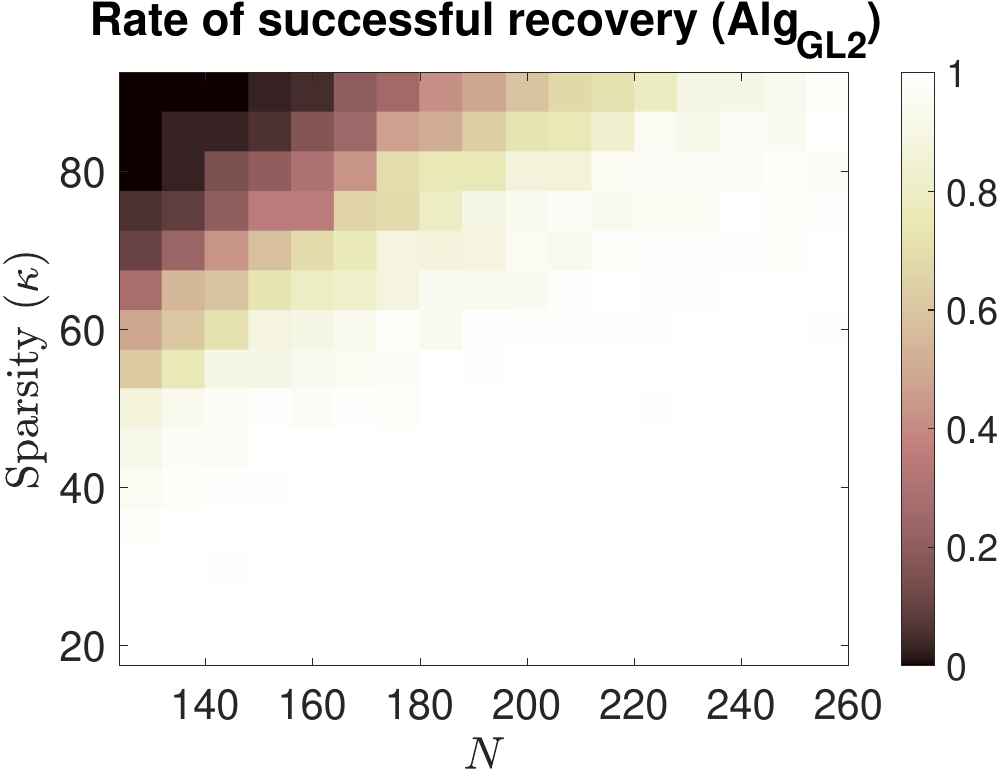}
\centerline{\footnotesize{(b)}}
\end{minipage}
\hfill
\begin{minipage}{0.3\linewidth}
\centering
\includegraphics[width=1.6in]{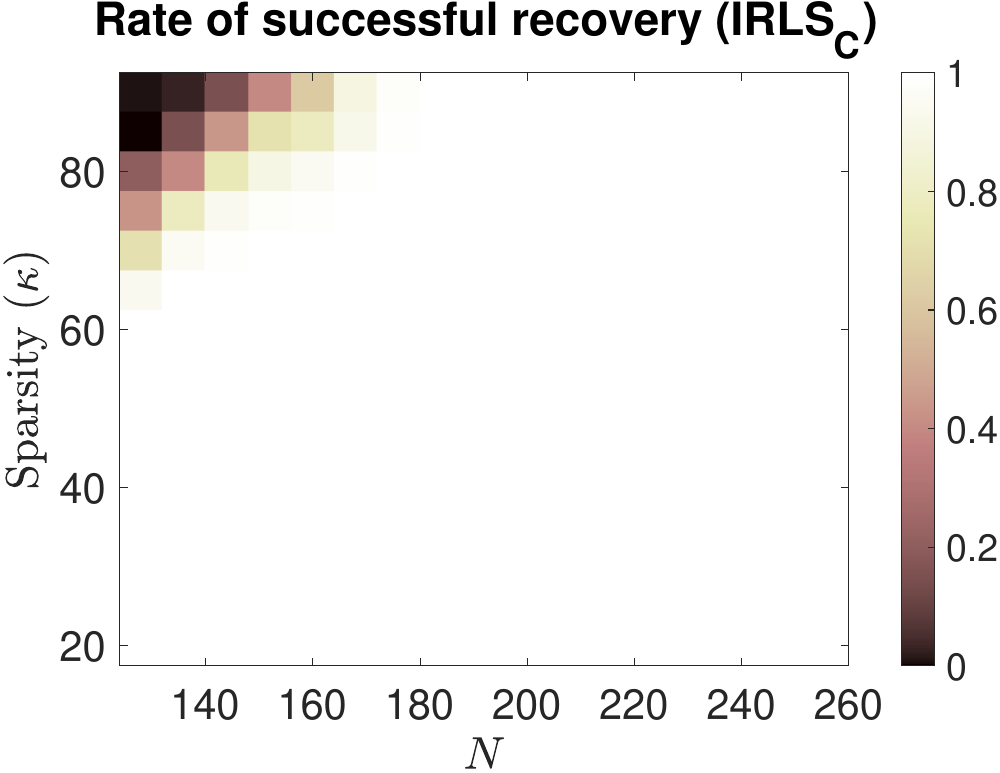}
\centerline{\footnotesize{(c)}}
\end{minipage}

\begin{minipage}{0.3\linewidth}
\centering
\includegraphics[width=1.6in]{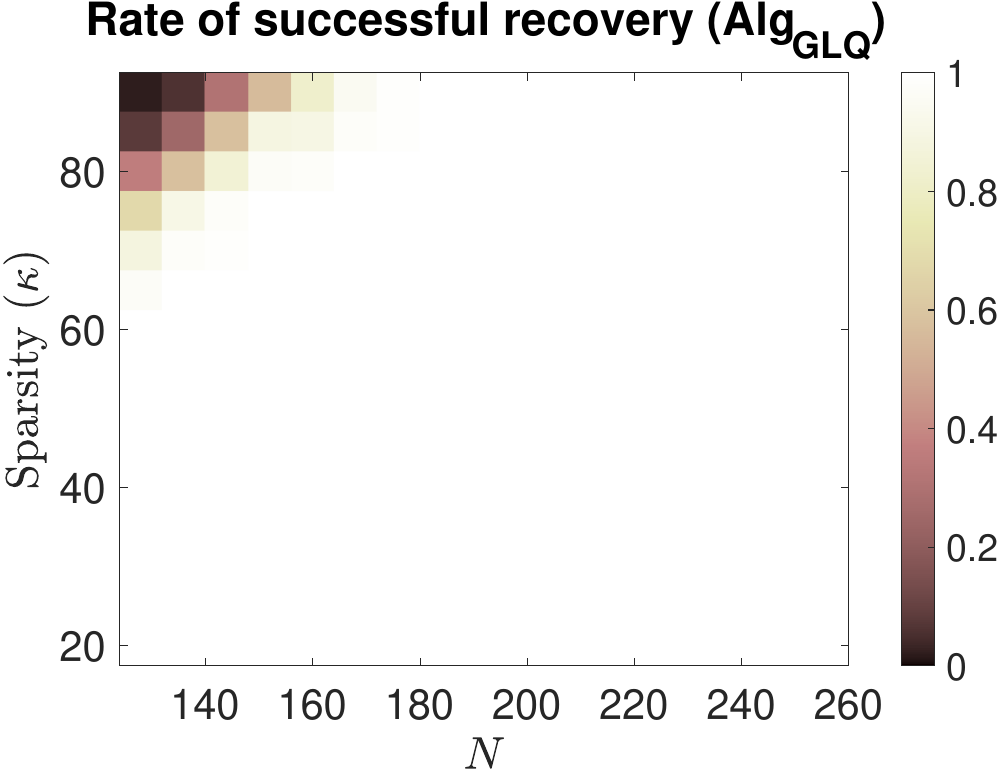}
\centerline{\footnotesize{(d)}}
\end{minipage}
\hfill
\begin{minipage}{0.3\linewidth}
\centering
\includegraphics[width=1.6in]{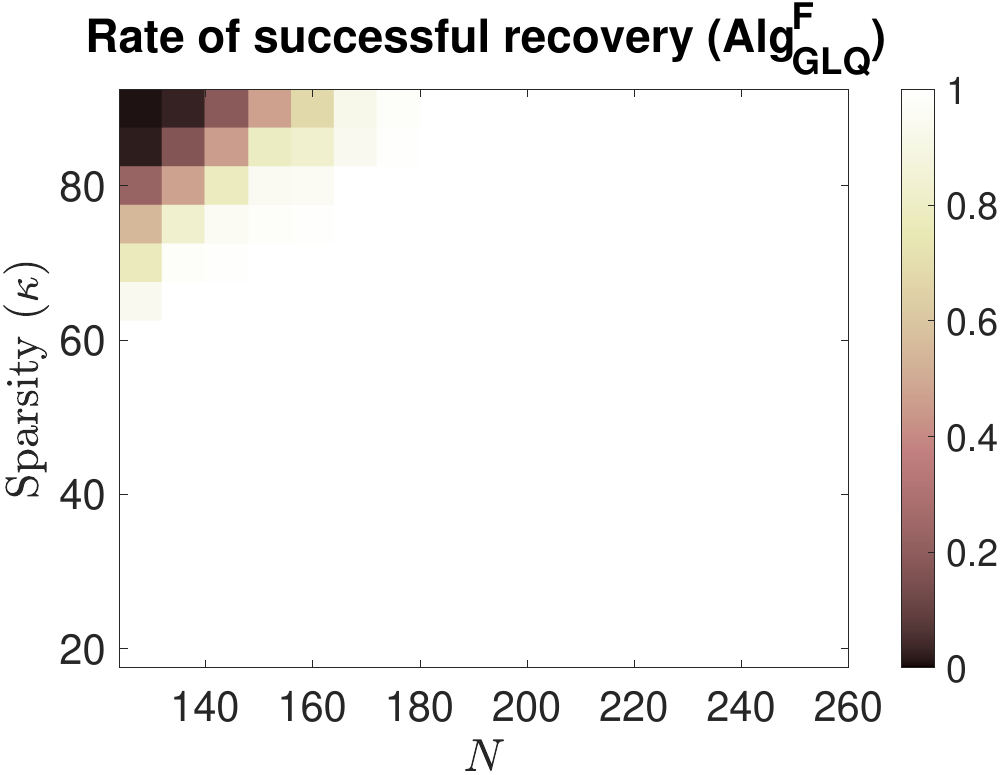}
\centerline{\footnotesize{(e)}}
\end{minipage}
\hfill
\begin{minipage}{0.3\linewidth}
\centering
\includegraphics[width=1.6in]{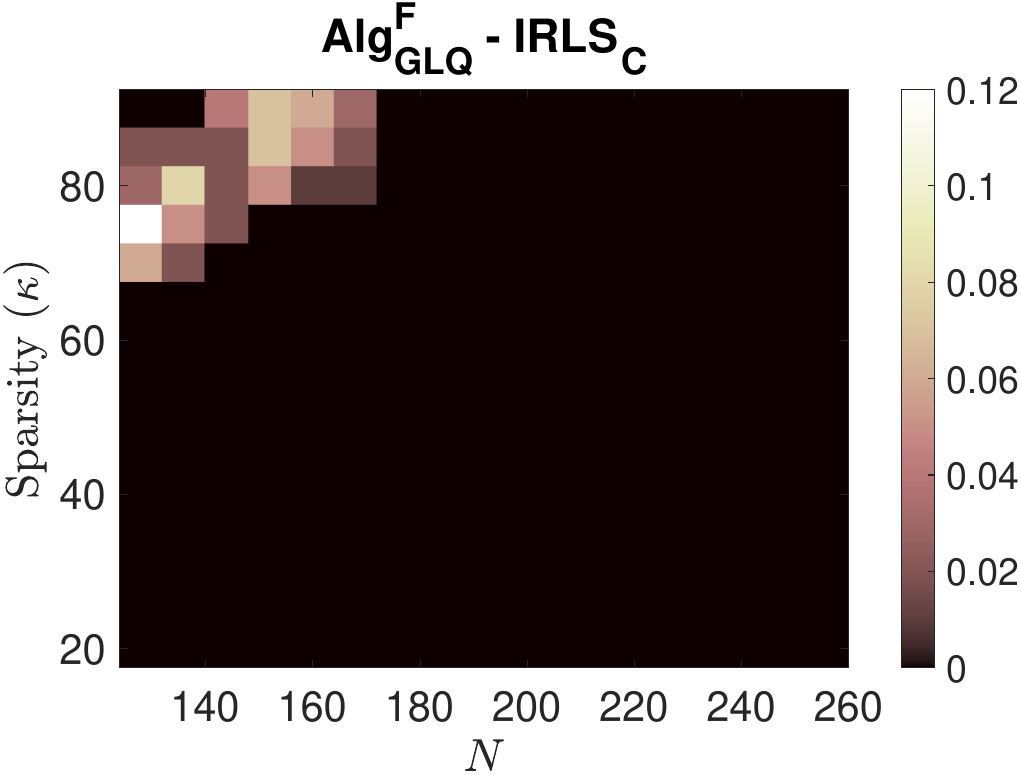}
\centerline{\footnotesize{(f)}}
\end{minipage}
\caption{The rate of successful recovery $\varrho_{ok}$ with respect to a series of sparsity level $\kappa$ and signal dimension $N$. These phase transition plots are generated with $J=100$ samples and with the setting $(\kappa, N, L=2N)$. }
\label{fig:test_phasetransitionX}
\end{figure*}

To investigate the impact of sparsity level $\kappa$ and signal dimension $N$ on the performance of the five algorithms: $\text{OMP}_C$,   $\text{IRLS}_C$, $\text{Alg}_{GL2}$, $\text{Alg}_{GLQ}$ and $\text{Alg}^F_{GLQ}$, we repeat the above experiments with $J=100$ samples and generate phase transition plots, as shown in Fig.~\ref{fig:test_phasetransitionX}. These experiments span signal dimensions $128 \leq N \leq 256$ and sparsity levels $20 \leq \kappa \leq 90$. The results demonstrate that the proposed $\text{Alg}_{GLQ}$ algorithm significantly outperforms the other four algorithms in terms of successful recovery rates. In addition, its accelerated version,  $\text{Alg}^F_{GLQ}$,  achieves a slightly higher successful recovery rate compared to $\text{IRLS}_C$, as shown in Fig.~\ref{fig:test_phasetransitionX} (f), which highlights the differences in successful recovery rates between $\text{Alg}^F_{GLQ}$ and $\text{IRLS}_C$.

\subsubsection{Effects of system matrix conditioning}
From the above, we note that the classical $\text{IRLS}_C$ and our proposed $\text{Alg}_{GLQ}$ and $\text{Alg}^F_{GLQ}$ yield almost identical performance for those systems $\mQ$ generated using $randn(N,L)$ and subsequently normalized. These matrices are well-conditioned. We note from \eqref{matrix-inv-LS-cla}, $\text{IRLS}_C$ may encounter numerical issues when $\mQ$ is ill-conditioned.

\begin{figure}[htb!]
\begin{minipage}{0.49\linewidth}
\centering
\includegraphics[width=2.2in]{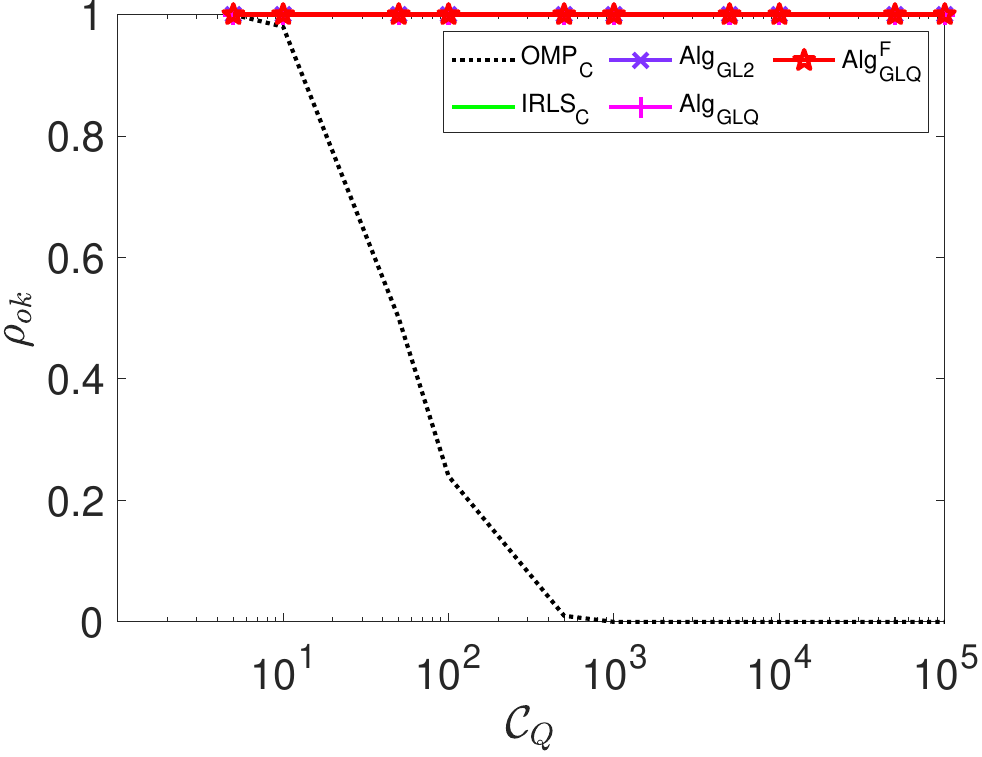}
\centerline{\footnotesize{(a)}}
\end{minipage}
\hfill
\begin{minipage}{0.49\linewidth}
\centering
\includegraphics[width=2.2in]{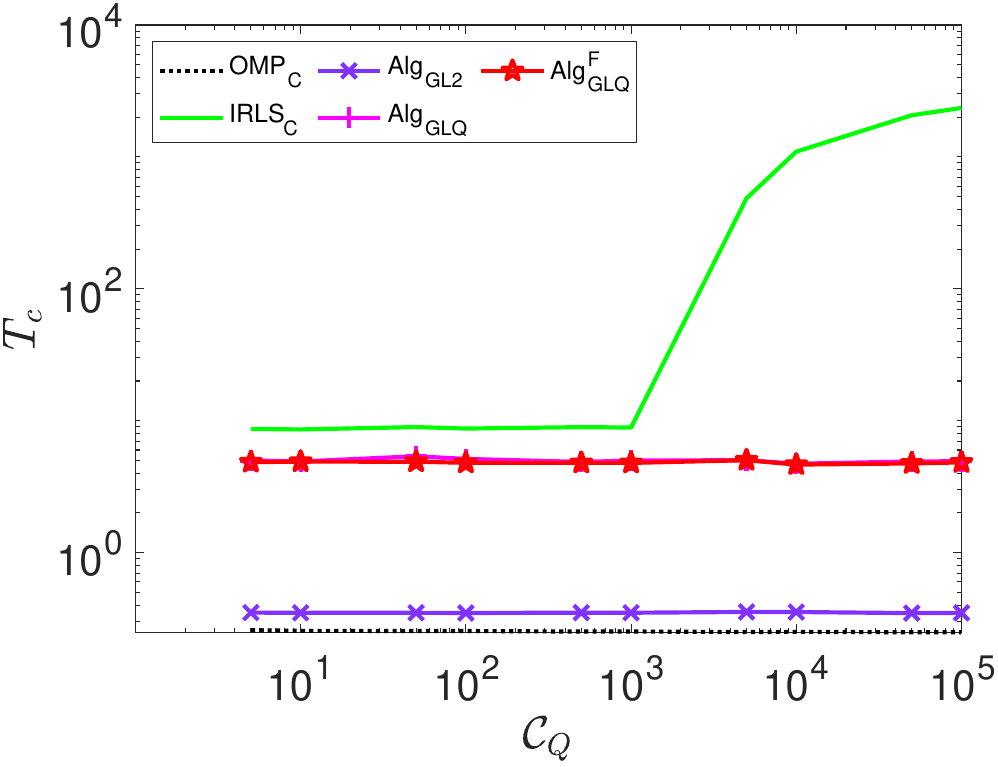}
\centerline{\footnotesize{(b)}}
\end{minipage}
\caption{The rate $\varrho_{ok}$ of successful recovery and the wall-clock time $T_c$  with respect to $\mathcal{C}_{\mQ}$ for each of the five algorithms for $\kappa=25, N=256, L=512$ and $J=100$. }
\label{fig:test_condQ_N256}
\end{figure}
\begin{figure}[htb!]
\begin{minipage}{0.49\linewidth}
\centering
\includegraphics[width=2.2in]{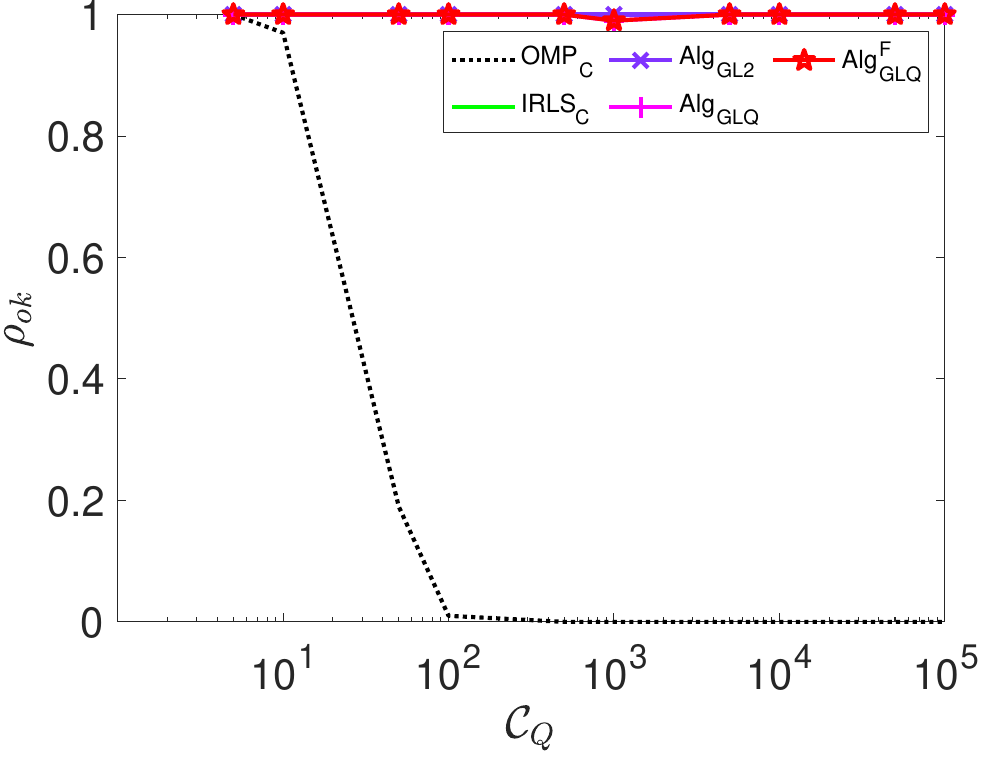}
\centerline{\footnotesize{(a)}}
\end{minipage}
\hfill
\begin{minipage}{0.49\linewidth}
\centering
\includegraphics[width=2.2in]{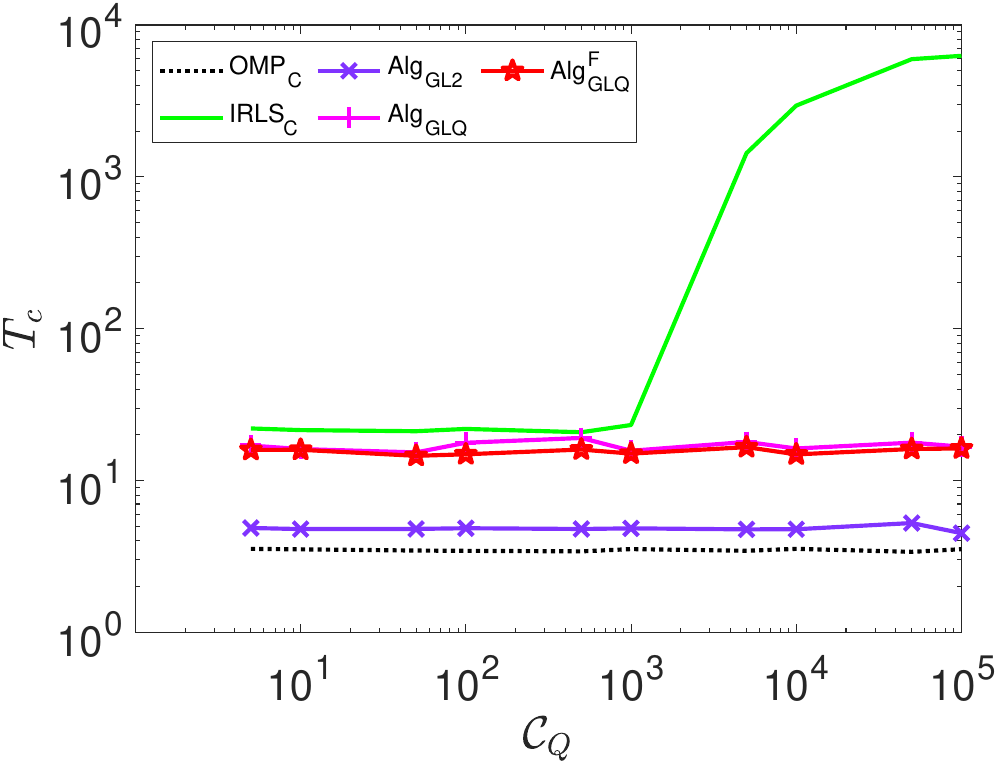}
\centerline{\footnotesize{(b)}}
\end{minipage}
\caption{The rate $\varrho_{ok}$ of successful recovery and the wall-clock time $T_c$  with respect to $\mathcal{C}_{\mQ}$ for each of the five algorithms for $\kappa=60, N=512, L=1024$ and $J=100$. }
\label{fig:test_condQ_N512}
\end{figure}

Let $\mathcal{C}_{\mQ}\triangleq \frac{\sigma_1}{\sigma_N}$ denote the condition number of $\mQ$, where $\sigma_1$ and $\sigma_N$ are the biggest and smallest of singular values of $\mQ \in \mathbb R^{N\times L}$. For a given pair $(N,L)$, we can generate a system matrix $\mQ$ in the same way as used in \cite{borgerding2017amp} for a specified $\mathcal{C}_{\mQ}$. 
We now consider the following settings: (i) $\kappa=25, N=256, L=512$ and $J=100$; and (ii) $\kappa=60, N=512, L=1024$ and $J=100$. Figs.~\ref{fig:test_condQ_N256}  and \ref{fig:test_condQ_N512} depict the statistics for $\varrho_{ok}$ and $T_c$ under each setting, respectively, where the system matrices $\mQ \in \mathbb R^{N\times L}$ are generated with  $\mathcal{C}_{\mQ}$ varies within
$[5~~  10~~  50~~  100~~  500~~  1,000~~  5,000~~  10,000~~  50,000~~  100,000].$

\begin{figure}[htb!]
\centering
\includegraphics[scale=0.34]{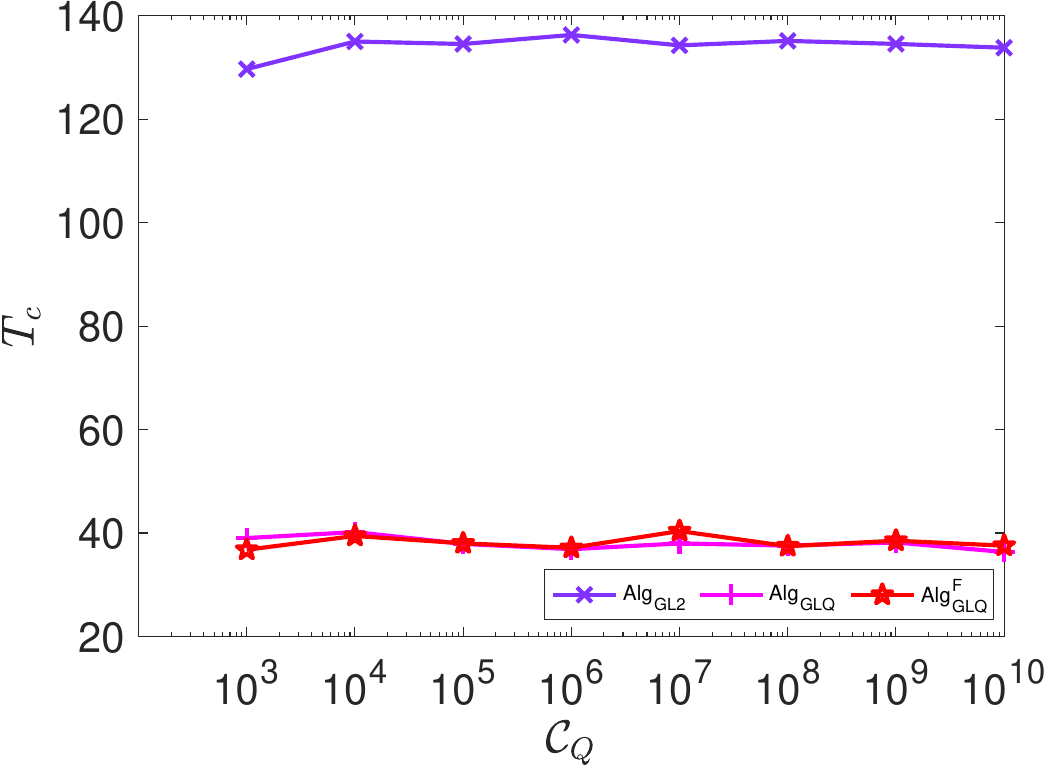}
\caption{The wall-clock time $T_c$ of the three algorithms  with respect to $\mathcal C_{\mQ}$ for $\kappa=500, N=2048, L=4096$ and $J=10$.}
\label{fig:test_condQ_N2048}
\end{figure}

It is observed that the $\varrho_{ok}$ of the classical $\text{OMP}_C$ is very sensitive to  $\mathcal{C}_{\mQ}$. This is not the case for our proposed $\text{Alg}_{GL2}$. Furthermore, 
1) the $T_c$ of $\text{IRLS}_C$ increases dramatically when $\mathcal{C}_{\mQ}$ exceeds $10^3$, reaching values hundreds of times greater than those of  $\text{Alg}_{GLQ}$ and $\text{Alg}^F_{GLQ}$. Indeed, when $\mathcal{C}_{\mQ}$ exceeds $10^6$, $\text{IRLS}_C$ encounters numerical issues and fails to work properly;
2) the condition number  $\mathcal{C}_{\mQ}$ has almost no effect on the performance of the three of our proposed algorithms in terms of $\varrho_{ok}$ and $T_c$. This is further confirmed under the setting $(\kappa, N, L)=(500, 2048, 4096)$ and $J=10$, where $\mathcal{C}_{\mQ}=10^m$ varies from $m=3, 4, 5, 6, 7, 8, 9, 10$. For all the proposed three algorithms, $\varrho_{ok}$ is constantly equal to one, while the Wall-clock time $T_c$ is presented in Fig.~\ref{fig:test_condQ_N2048}.

In summary,
\begin{itemize} \item for well-conditioned systems,  our proposed algorithms $\text{Alg}_{GL1}, \text{Alg}^F_{GL1}$, $\text{Alg}_{GLQ}$ and $\text{Alg}^F_{GLQ}$ achieve the same or even better sparse recovery performance compared to the state-of-the-art performance algorithm  $\text{IRLS}_C$. Additionally, the proposed $\text{Alg}_{GL2}$ 
significantly outperforms the classical $\text{OMP}_C$ and $\text{BP}_C$ in terms of sparse recovery, with computational efficiency comparable to   $\text{OMP}_C$;
\item for ill-conditioned systems, the performance of all classical algorithms degrades dramatically in terms of either sparse recovery accuracy or computational efficiency, or both. In contrast, the three proposed algorithms $\text{Alg}_{GL2}$, $\text{Alg}_{GLQ}$ and $\text{Alg}^F_{GLQ}$ can work properly even for systems with $\mathcal{C}_{\mQ}$ up to $10^{10}$ and large-scale dimensions.
\end{itemize}


\subsection{Robustness against low-rank disturbance}
Next, we evaluate the performance of our proposed algorithms under the noisy signal model introduced in Section~\ref{sec:robust}, with a particular focus on comparing them to $\text{IRLS}_C$. 
\subsubsection{Synthetic data}
\begin{figure}[htb!]
\begin{minipage}{0.49\linewidth}
\centering
\includegraphics[width=2.2in]{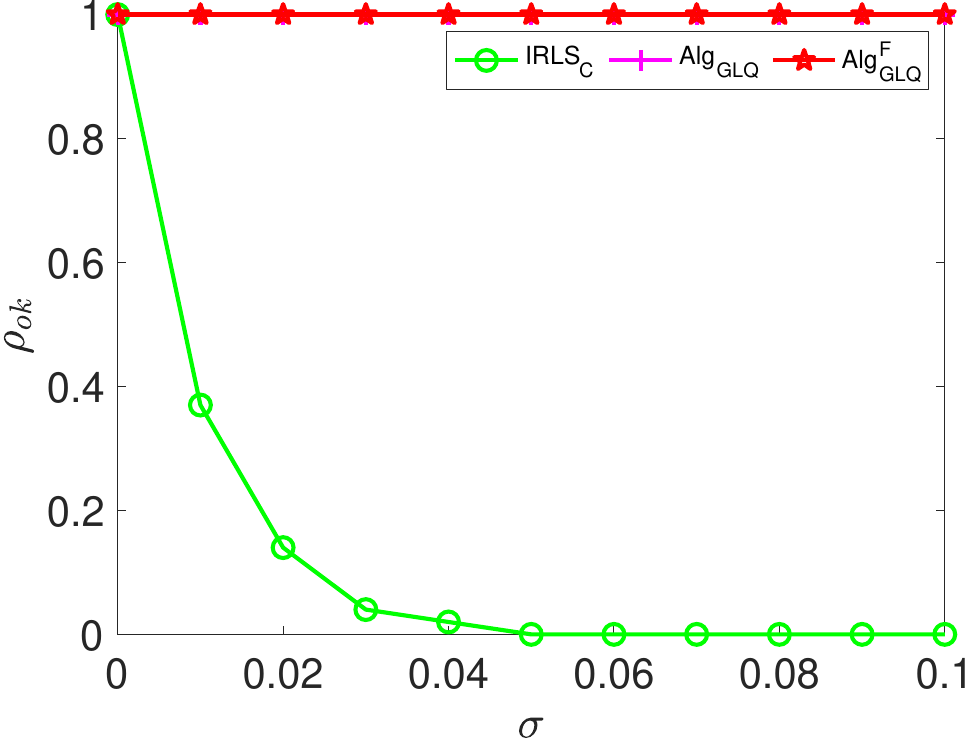}
\centerline{\footnotesize{(a)}}
\end{minipage}
\hfill
\begin{minipage}{0.49\linewidth}
\centering
\includegraphics[width=2.2in]{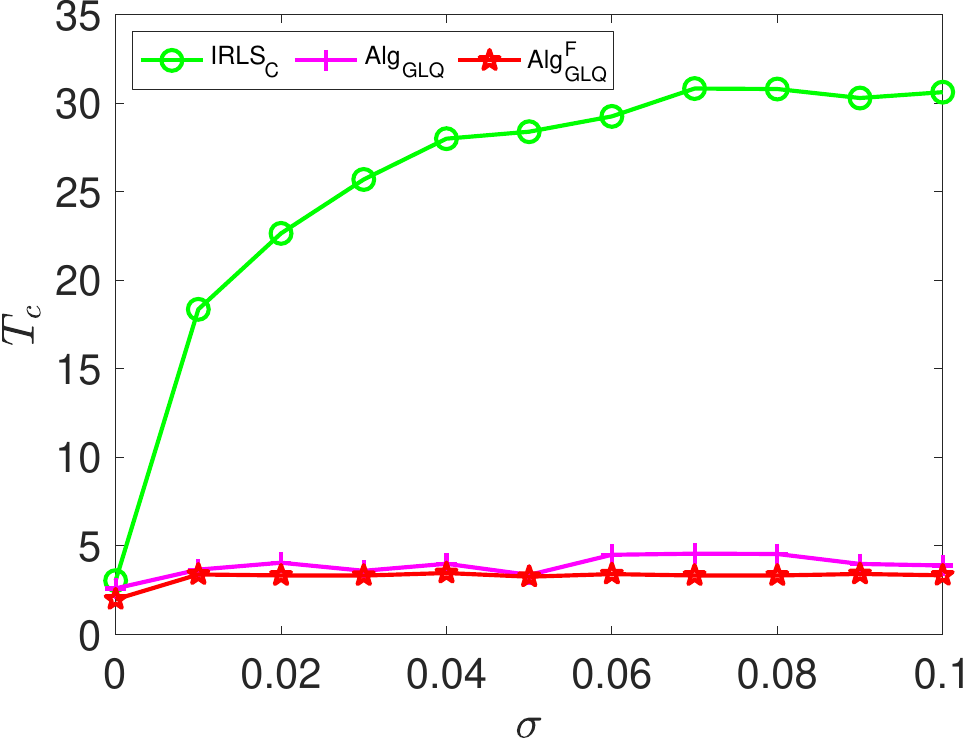}
\centerline{\footnotesize{(b)}}
\end{minipage}
\caption{The rate $\varrho_{ok}$ of successful recovery and the wall-clock time $T_c$  with respect to different noise variance $\sigma$ for each of the three algorithms for $N=128, L=256,J=100$. 
} \label{fig_noise_N128}
\end{figure}

We repeat the previous experiments with parameters $N = 128, L = 256, J = 100$, and $\kappa = 30$. We set $N_e=20$ and construct $\mGamma$ using the first $N_e$ left singular vectors of an $N\times N$ Gaussian random matrix with entries drawn from $\mathcal{N}(0,1)$. The coefficient $\valpha^\star$ in~\eqref{restart-3} is then generated as $\valpha^\star = \sigma \valpha_0$, where $\valpha_0$ is a normalized Gaussian random vector with entries following $\mathcal{N}(0,1)$. Here, our primary focus is on comparing the performance of our proposed $\text{Alg}_{GLQ}$ and $\text{Alg}^F_{GLQ}$ algorithms against  $\text{IRLS}_{C}$. As illustrated in Fig.~\ref{fig_noise_N128}, our proposed methods demonstrate significant advantages over $\text{IRLS}_{C}$ algorithm in terms of both successful recovery rate and computational efficiency. Notably, as the noise level $\sigma$ increases to $0.05$, $\text{IRLS}_{C}$ completely fails to reconstruct the $J=100$ signals, whereas the proposed $\text{Alg}_{GLQ}$ and $\text{Alg}^F_{GLQ}$ maintain perfect recovery performance across all the $J=100$ signals.

\subsubsection{Image transmission}
\begin{figure}[t]
\begin{minipage}{0.01\linewidth}  
    \rotatebox{90}{\footnotesize MRI1}  
\end{minipage}
\begin{minipage}{0.11\linewidth}
\centerline{\footnotesize{Clean}}
\centering
\includegraphics[width=0.7in]{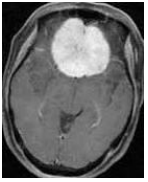}
\end{minipage}
\hfill
\begin{minipage}{0.11\linewidth}
\centerline{\footnotesize{Noisy}}
\centering
\includegraphics[width=0.7in]{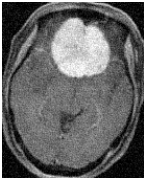}
\end{minipage}
\hfill
\begin{minipage}{0.11\linewidth}
\centerline{\footnotesize{$\text{OMP}_C$}}
\centering
\includegraphics[width=0.7in]{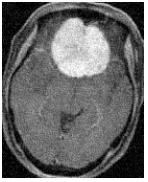}
\end{minipage}
\hfill
\begin{minipage}{0.11\linewidth}
\centerline{\footnotesize{$\text{IRLS}_C$}}
\centering
\includegraphics[width=0.7in]{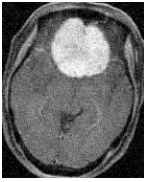}
\end{minipage}
\hfill
\begin{minipage}{0.11\linewidth}
\centerline{\footnotesize{$\text{Alg}_{GL1}$}}
\centering
\includegraphics[width=0.7in]{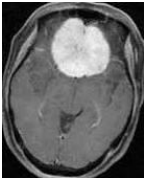}
\end{minipage}
\hfill
\begin{minipage}{0.11\linewidth}
\centerline{\footnotesize{$\text{Alg}_{GL1}^F$}}
\centering
\includegraphics[width=0.7in]{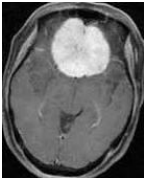}
\end{minipage}
\hfill
\begin{minipage}{0.11\linewidth}
\centerline{\footnotesize{$\text{Alg}_{GLQ}$}}
\centering
\includegraphics[width=0.7in]{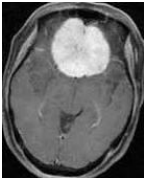}
\end{minipage}
\hfill
\begin{minipage}{0.11\linewidth}
\centerline{\footnotesize{$\text{Alg}^F_{GLQ}$}}
\centering
\includegraphics[width=0.7in]{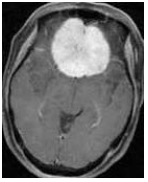}
\end{minipage}
\\
\begin{minipage}{0.01\linewidth}  
    \rotatebox{90}{\footnotesize MRI2}  
\end{minipage}
\begin{minipage}{0.11\linewidth}
\centering
\includegraphics[width=0.7in]{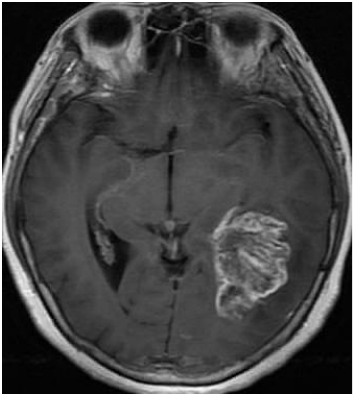}
\end{minipage}
\hfill
\begin{minipage}{0.11\linewidth}
\centering
\includegraphics[width=0.7in]{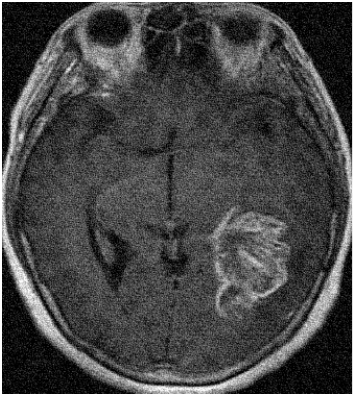}
\end{minipage}
\hfill
\begin{minipage}{0.11\linewidth}
\centering
\includegraphics[width=0.7in]{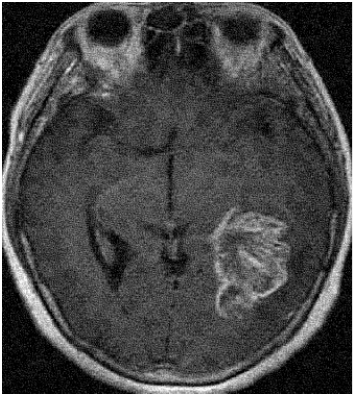}
\end{minipage}
\hfill
\begin{minipage}{0.11\linewidth}
\centering
\includegraphics[width=0.7in]{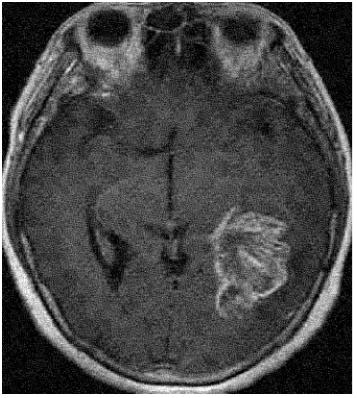}
\end{minipage}
\hfill
\begin{minipage}{0.11\linewidth}
\centering
\includegraphics[width=0.7in]{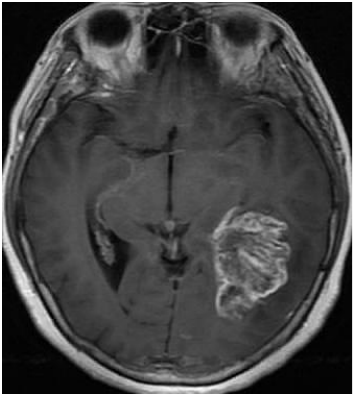}
\end{minipage}
\hfill
\begin{minipage}{0.11\linewidth}
\centering
\includegraphics[width=0.7in]{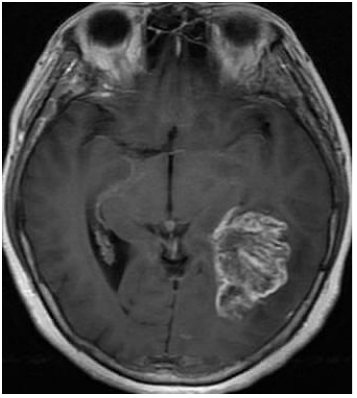}
\end{minipage}
\hfill
\begin{minipage}{0.11\linewidth}
\centering
\includegraphics[width=0.7in]{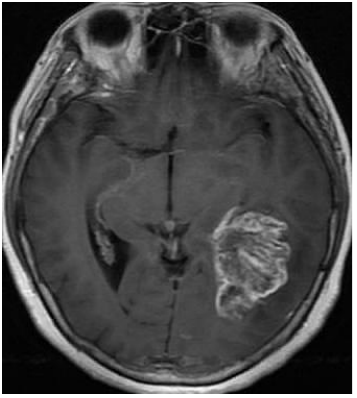}
\end{minipage}
\hfill
\begin{minipage}{0.11\linewidth}
\centering
\includegraphics[width=0.7in]{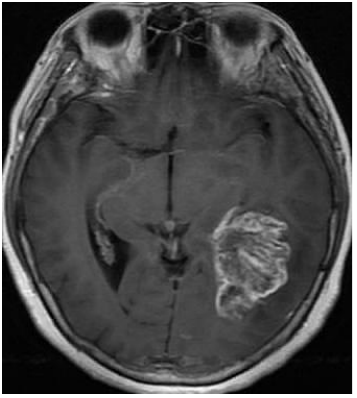}
\end{minipage}
\\
\caption{Image reconstruction by the six algorithms: $\text{OMP}_C$, $\text{IRLS}_C$, $\text{Alg}_{GL1}$, $\text{Alg}^F_{GL1}$, $\text{Alg}_{GLQ}$, and $\text{Alg}^F_{GLQ}$.}
\label{fig:test_images}
\end{figure}

 We consider an image transmission system affected by a low-rank interference signal $\ve$ modeled as in~\eqref{restart-3}, with $\mGamma \in \mathbb R^{64\times 5}$ assumed to be known. The clean images intended for transmission are shown in the first column of Fig.~\ref{fig:test_images}. Prior to transmission, each clean image is divided into patches of size $8\times 8$, resulting in an $N \times J$ data matrix with $N=64$ and $J$ corresponding to the number of patches. After transmission, the received noisy images are presented in the second column of Fig.~\ref{fig:test_images}. 

Given the noisy images, we reconstruct the clean images using six different algorithms: $\text{OMP}_C$, $\text{IRLS}_C$, $\text{Alg}_{GL1}$, $\text{Alg}^F_{GL1}$, $\text{Alg}_{GLQ}$, and $\text{Alg}^F_{GLQ}$. In the reconstruction process, we employ an overcomplete discrete cosine transform (DCT) dictionary $\mQ$ of size $64 \times 144$, i.e., $N=64, L=144$. The sparsity level $\kappa$ is set differently for each image, with specific values provided in Table~\ref{table-denoising}. 
The corresponding image sizes are also listed in Table~\ref{table-denoising}. The reconstructed images obtained from the six algorithms are displayed in Fig.~\ref{fig:test_images}. Table~\ref{table-denoising} further presents the peak signal-to-noise ratio (PSNR, in dB) for each method. The results demonstrate that our four proposed algorithms  achieve significantly higher PSNR values compared to $\text{OMP}_C$ and $\text{IRLS}_C$, highlighting their superior performance in reconstructing the clean images.

\begin{table}[thb!]\caption{Statistics of PSNR (in dB) for the six algorithms in image reconstruction.}\label{table-denoising}
\centering
\smaller
\setlength{\tabcolsep}{3pt}
\renewcommand{\arraystretch}{1.05}
\resizebox{\linewidth}{!}{%
\begin{tabular}{|c|c|c||c|c|c|c|c|c|c|}
\hline \cline{1-10}  \hline \cline{1-10}
 & Size & $\kappa$&  Noisy &$\text{OMP}_C$ & $\text{IRLS}_C$&$\text{Alg}_{GL1}$&$\text{Alg}^F_{GL1}$&$\text{Alg}_{GLQ}$&$\text{Alg}^F_{GLQ}$\\ \cline{1-6}  \hline \cline{1-10}
MRI1&$168\times 136$ & 20   & 24.42 & 25.03&25.13 & 50.45    &49.72    &49.94    &49.90       \\ \cline{1-10}
MRI2&$376\times 336$ & 12   & 20.47 & 22.05&22.22 & 55.22    &55.01    &55.01    &54.83       \\ \cline{1-10}
\end{tabular}
}
\end{table}

\section{Concluding remarks}\label{sec-6}




In this paper, we have investigated the sparse signal recovery problems. Our study has led to several key findings and contributions that enhance the field of sparse signal recovery.
Based on a proposed system model, we have developed a novel approach leading to a class of greedy algorithms, including
\begin{itemize}
    \item three innovative greedy algorithms, $\text{Alg}_{GL2}$, $\text{Alg}_{GL1}$ and $\text{Alg}_{GLQ}$, which significantly improve the accuracy of classical methods such as OMP and BP; 
    \item  two fast algorithms, $\text{Alg}^F_{GL1}$ and $\text{Alg}^F_{GLQ}$, designed to accelerate $\text{Alg}_{GL1}$ and $\text{Alg}_{GLQ}$, achieving performance that even surpasses {\it state-of-the-art} sparse recovery methods.
\end{itemize}
 Furthermore,  the five proposed algorithms demonstrate excellent robustness against instability caused by the poor numerical properties of the system matrix $\mQ$ and measurement interferences -- issues that significantly affect classical sparse recovery algorithms.  
    




     Future work will focus on further enhancing the algorithms' efficiency and exploring their applications in more complex high-dimensional data analysis tasks, such as hyperspectral imaging, seismic data processing, and wireless communication. In conclusion, the proposed greedy algorithms represent a significant advancement in sparse recovery techniques. By combining theoretical rigor with practical efficiency, our work paves the way for future innovations and applications in the field of signal processing and beyond.

\bibliographystyle{unsrt}
\bibliography{reference}


\end{document}